\documentclass[english]{article}
\RequirePackage[OT1]{fontenc}
\RequirePackage{graphicx}
\RequirePackage{amsthm}
\RequirePackage{amsmath}
\RequirePackage[authoryear]{natbib}
\RequirePackage[colorlinks,citecolor=blue,urlcolor=blue]{hyperref}
\usepackage{algorithm,algorithmic}
\usepackage{amssymb}
\usepackage{footnote}
\usepackage{subfig}
\usepackage{array}
\usepackage{setspace}
\usepackage{comment}
 \usepackage{tikz}
\usetikzlibrary{decorations.pathreplacing}
\theoremstyle{plain}

\numberwithin{equation}{section}
\numberwithin{figure}{section}

\graphicspath{ {./plotsforarxiv/} }

\makeatletter
\newcommand{\indep}{\perp \!\!\! \perp}

\@ifundefined{showcaptionsetup}{}{%
	\PassOptionsToPackage{caption=false}{subfig}}
\makeatother

\begin{document}
	\title{Causal Duration Analysis with  Diff-in-Diff\thanks{We thank Arnim Seidlitz for excellent research assistance. We  thank seminar participants at Brown University, The University of Surrey, Goethe University, as well as those who attended the presentations of this paper at the European Winter Meeting of the Econometric Society 2023, the 2025 Econometric Society World Congress, and the 2024 Bristol Econometric Study Group, for helpful comments and feedback. We would also like to thank Guido Imbens, Xavier D'Haultfoeuille, anonymous referees, and colleagues at UCL for their insights, particularly Christian Dustmann, Dennis Kristensen, Liyang Sun, and Andrei Zeleneev, and others who attended internal presentations of this work as well as a number of anonymous referees.\\ Replication code and a STATA command implementing our main procedures can be found along with instructions at \href{https://github.com/ben-deaner-teaching/Duration-DiD/tree/main}{https://github.com/ben-deaner-teaching/Duration-DiD/tree/main}}}
	\author{Ben Deaner\thanks{b.deaner@ucl.ac.uk} \,\,and Hyejin Ku\thanks{h.ku@ucl.ac.uk}\\ University College London }
	\maketitle
	\begin{abstract}
		
		\begin{center}
			
		\end{center}
		
		In economic program evaluation, it is common to obtain panel data in which outcomes are indicators that an individual has reached an absorbing state. For example, they may indicate whether an  individual has exited a period of unemployment, passed an exam, left a marriage, or had their parole revoked. The parallel trends assumption that underpins difference-in-differences generally fails in such settings. We suggest identifying conditions similar to those of difference-in-differences, 
		but which apply to hazard rates rather than mean outcomes. 
		These alternative assumptions motivate estimators that retain the simplicity and transparency of standard diff-in-diff, and corresponding specification tests. Our approach can be adapted to include general linear restrictions between the hazard rates of different groups, motivating duration analogues of the triple differences and synthetic control methods. We apply our procedures to examine the impact of a policy that increased the generosity of unemployment benefits, using a cross-cohort comparison.
	\end{abstract}
	
	\newpage
	
	Many important topics in applied economics involve durations. To name a few, the impact of unemployment insurance on the length of unemployment spells (\cite{Katz1990}, \cite{Hunt1995}, \cite{LALIVE2006}, \cite{Lalive2007}, \cite{Card2007}, \cite{Chetty2008}, \cite{Schmieder2012}, \cite{Schmieder2016}, \cite{Lichter2021}, and others), the effect of divorce laws on marriage duration (\cite{Friedberg1998a}, \cite{Gruber2004}, \cite{Wolfers2006}), the strength of residency rules on the rate at which refugees pass language tests (\cite{Arendt}), the impact of insurance schemes on fertility rates (\cite{Lambert2016}), mandatory retirement rules on rates of retirement (\cite{Ashenfelter2002}), and the consequences of criminal justice policies on the rate of recidivism or probation revocation (\cite{Schmidt1989}, \cite{Bhuller2020}, \cite{Rose2021}). In settings like these, available data may consist of panels or repeated cross-sections in which the outcome is a binary indicator that an individual has entered an absorbing state. For example, an indicator that an individual has exited unemployment by a particular date. Difference-in-differences is a popular tool for policy-evaluation, but when the data take the form just described, the parallel trends assumption generally fails.
	
	To fix ideas, consider the case in which the outcome indicates exit from unemployment. Suppose that some individuals receive an increase in unemployment benefits at a particular point in time while others do not. The foundational assumption of diff-in-diff is that, absent the policy of interest, the difference in mean outcomes between the treated and untreated groups would remain fixed. If a sufficiently large share of individuals eventually exit unemployment, then mean outcomes will tend to converge over time, even absent any treatment effect.\footnote{Apart from in the special case in which mean outcomes are identical between the two groups.}  This entails a failure of parallel trends and may result in severely biased and inconsistent estimates of treatment effects.
	
	It is well-known that the parallel trends assumption is generally incompatible with settings in which outcomes are discrete or truncated (see \cite{Wooldridge2023}). What makes this problem particularly acute in duration settings is that the resulting bias is systematic. Suppose a higher proportion of individuals in the treated group are unemployed just prior to treatment. Then under a counterfactual in which no one is treated, the gap in unemployment rates between the treated and untreated groups would tend to narrow over the post-treatment period. Standard diff-in-diff estimates attribute this narrowing to the effect of treatment, leading to over-estimation of the effect of treatment on job-finding. By contrast, in cases in which the binary outcomes reflect say, choices among discrete alternatives, it is less clear that the misspecification of the model would systematically bias the estimates in any particular direction.

	In response, we consider alternative
	approaches based on insights from duration analysis. Rather than assume
	parallel trends between mean potential outcomes, we impose restrictions on the dynamics of group-specific counterfactual hazard rates. For example, we may assume that changes in the hazard rate over any time interval are equal across groups, or that changes in the log hazard rate are equal across groups. These conditions are consistent with the convergence of counterfactual mean outcomes over time, and thus the convergence does not imply inconsistency of the corresponding causal estimates. Our assumptions motivate a simple fix. Instead of performing diff-in-diff or related methods using mean outcomes, we apply the same procedures using a particular transformation of the mean outcomes, similar to \cite{Wooldridge2023}. We sidestep any estimation of the hazard rates themselves. Our methods are nonparametric in that we avoid specifying a parametric model for group-specific hazard rates.
	
	Diagnostics like tests for pre-treatment parallel trends can be adapted straight-forwardly to our setting. Where standard tests use mean outcomes, we use transformed means. Similarly to vanilla diff-in-diff, one can perform informal visual inspections to assess whether parallel trends in these objects holds to a close approximation in the pre-treatment period.
	
	Adjusting for covariates can be important for credibly identifying causal effects in difference-in-differences. If there is covariate imbalance between treatment and control groups in diff-in-diff, then parallel trends may fail, even if it holds within each covariate stratum (say, within each demographic subgroup). We suggest a nonparametric covariate balancing approach, similar to \cite{Abadie2005}. We also provide an alternative semiparametric method of covariate adjustment in Section A.4 of the supplementary appendix.
	
	Our approach extends beyond the assumption of a fixed level difference between the hazard rates of different groups or their logarithms. We can accommodate any time-invariant linear relationship between counterfactual hazard rates. In the case of multiple untreated groups we can obtain a  duration analogue of the triple differences estimator and of the synthetic control method. As with our duration diff-in-diff approach, estimation differs from standard approaches only in the use of transformed mean outcomes.  In Section A.3 of the supplementary appendix, we extend our main identification results to settings with staggered adoption and suggest simple estimators.
	
	We apply our methods to the setting of \cite{LALIVE2006}. The authors in that study evaluate the impact of a policy that increased the generosity of unemployment insurance benefits for unemployed Austrian workers. The authors identify causal effects by exploiting the presence of individuals who were ineligible for the benefit changes. They estimate a flexible parametric duration model and from their estimates they recover causal effects.
	
	In contrast, we identify causal effects using a cross-cohort comparison. We employ our methods in order to estimate the impact of an extension to the potential benefit duration (PBD). We adjust for the calendar date at which an unemployment spell begins using our covariate balancing strategy. Thus our estimates are robust to differential trends in job-seeking between individuals who become unemployed in different parts of the year. We obtain similar results to the authors of the original study, however we do so while avoiding estimation and specification of the hazard function, and without the need for numerical optimization of a likelihood. Moreover, the transparency of our approach allows us to both visually and formally assess whether there are parallel trends over the pre-treatment period. We find a statistically significant positive impact of the PBD extension on unemployment rates with strongly positive estimates shortly following treatment which then taper off. We are unable to reject parallel trends in hazard rates even at the $60\%$ level. 
	
	We evaluate the finite-sample performance of our methods in a simulation study. The results also demonstrate the potential for standard diff-in-diff to produce severely misleading estimates in duration settings. The simulation results are available in Section C of the supplementary appendices.

	
	\subsection*{Related Literature}
	
	We are not the first to suggest an extension of differences-in-differences to duration settings. Proportional hazard models (\cite{Cox1972}) with a diff-in-diff-type linear index are considered by \cite{Hunt1995},  \cite{Wu2022}, \cite{LALIVE2006}, and \cite{Marinescu2009}. \cite{Wu2022} show that parallel trends cannot hold when the data are generated by the proportional hazards model. \cite{Hunt1995} suggests an approximate partial likelihood estimation procedure for the coefficients of the linear index in a proportional hazard diff-in-diff specification, \cite{LALIVE2006} and \cite{Marinescu2009} employ maximum likelihood estimation. \cite{Wu2022} consider proportional hazard estimation in the two-period case. In both \cite{Wu2022} and \cite{Hunt1995},  interest is in the estimation of the coefficients on the binary indicators in the linear index, rather than average treatment effects.
	
	Also related to our approach is the literature on non-linear difference-in-differences. These papers specify  a generalized linear model (GLM) in which the outcome is a non-linear transformation of a linear index with a diff-in-diff form. A number of empirical papers estimate the coefficients
	in a GLM diff-in-diff model, for example \cite{Gruber1994} and \cite{Feldstein1996}. \cite{Puhani2011} considers the interpretation of the coefficients in these models. \cite{Ashenfelter2004} allows the link function to be estimated using sufficiently rich pre-treatment data. \cite{Athey2006} provide a method that resembles difference-in-differences but avoids functional form restrictions.
	
	Motivation for GLM diff-in-diff is
	discussed in \cite{Blundell2004}, \cite{Blundell2009}, \cite{Lechner2011},  and most thoroughly in  \cite{Wooldridge2023}. Similar to the present work, the strategy in \cite{Blundell2004} and \cite{Wooldridge2023} is to perform diff-in-diff on  transformed mean outcomes. Inverting the transformation then recovers counterfactual mean outcomes. Their specifications are motivated by latent variable models for the discrete outcomes. In our work the form of the transformation follows from the duration structure of the data rather than a distributional assumption, effectively providing a micro-foundation for a special case of the approach in \cite{Wooldridge2023}. For further comparison to \cite{Wooldridge2023} see Section 2, Remark 4.

	Our work extends the ever-growing literature on difference-in-differences. For recent surveys see for example \cite{Roth2023} and \cite{Chaisemartin2023}. More general works that consider the failure of parallel trends under alternative functional forms include \cite{Roth} and \cite{Rambachan2023}. 
	
	In our empirical application we employ a cross-cohort comparison similar to \cite{Berg2020}, who employ an identification strategy based on regression discontinuity design for nonparametrically estimated hazard rates.
	
	There is a sizable literature on causal analysis using duration data.   See e.g.,  \cite{Abbring2007} for a survey and \cite{Abbring2003} for seminal work on this topic. \cite{Abbring} identify and estimate causal effects from duration data using instrumental variables. \cite{Vikstroem2017} use a dynamic inverse propensity score weighting to estimate treatment effects under unconfoundedness. \cite{Berg} use an inverse propensity score weighting to identify causal objects under a sequential unconfoundedness condition. \cite{Cui2023} use random forests to estimate conditional average treatment effects under a conditional ignorability assumption.

One can also apply difference-in-differences using durations themselves as outcomes, as in \cite{Lichter2021}. In the unemployment example, one would compare mean unemployment durations between those who became unemployed before and after the reform, and examine how this difference varies with eligibility. This approach differs from ours in applicability and in the causal objects it identifies.
Unlike our approach, it requires variation in unemployment spell start dates and censoring corrections if some spell lengths are discretized or incomplete. Moreover, if the policy applies to ongoing spells, then eligible individuals who became unemployed before the reform include some who are treated, and dropping these individuals introduces selection bias.
The two approaches also identify distinct causal objects. If the policy affects not only time spent unemployed but also who becomes unemployed, this compositional change is an additional channel through which the policy impacts mean durations. By contrast, we follow a fixed set of individuals who were unemployed prior to the reform and thus isolate their behavioral response.
	
	\section{Motivation and Background}
	
	Consider a sample of individuals indexed by $i=1,...,n$ observed at times $t=1,...,T$. Each individual undergoes a `spell', for example a period of unemployment, a marriage, or enrolment in a course of study. Let $Y_{i,t}$ be a binary indicator that individual $i$'s spell has ended by time $t$. We allow for the possibility that some spells have already ended by time $t=1$, but no individual begins a spell after this time. Each individual belongs to a group $G_i$ where group membership is constant over time. An intervention occurs at some time strictly after $t^{*}$ and impacts only individuals in Group $1$. We wish to assess the impact of this intervention on their outcomes $Y_{i,t}$.
	
It is sometimes useful to distinguish between the observation time $t$ and the calendar date to which $Y_{i,t}$ corresponds, which we denote $t(i)$. In our application to unemployment insurance, individuals enter unemployment on various calendar dates and are treated a fixed number of days into their spell. Thus the time $t$ is defined relative to the start of an individual's spell. The calendar date $t(i)$ is then equal to $t+s_i$, where $s_i$ is individual $i$'s unemployment start date. In other settings treatment (for example a divorce law reform) may occur on a fixed calendar date, in which case $t(i)=t$ for all $i$. 


	
	
	In order to define causal effects of the treatment, we consider a
	counterfactual in which there is no intervention on Group $1$.
	We denote by $Y_{i,t}^{(0)}$ the outcome under this counterfactual
	at time $t$ for individual $i$. We sometimes refer to $Y_{i,t}^{(0)}$
	as an `untreated potential outcome'.\footnote{Note that $Y_{i,t}^{(0)}$ differs from the
	standard definition of untreated potential outcomes in that the counterfactual is defined in terms
	of an intervention on Group $1$ rather than an individual-level
	treatment.}  We use the superscript `$(0)$' to indicate counterfactual values throughout this work. 
	
Our primary object of interest is the time-$t$ average treatment
	effect for individuals in the treated group.
	This is the average difference between the outcome for a randomly sampled individual
	in the treated Group $1$ at time $t$, and that individual's outcome in the counterfactual
	world in which there is no intervention. This is defined as follows:
	\[
	\tau_{t}:=E[Y_{i,t}-Y_{i,t}^{(0)}|G_i=1].
	\]
	Note that if all spells begin at $t=1$, then $E[Y_{i,t}^{(0)}|G_i=1]$ is the counterfactual cumulative distribution function (CDF) of the durations of individuals in Group $1$  evaluated at $t$. Therefore, if treatment effects are identified, then so too are the values of the counterfactual CDF of durations evaluated at each discrete time increment. Using terminology from duration analysis, $1-E[Y_{i,t}^{(0)}|G_i=k]$ is the counterfactual \textit{survival function} at time $t$ and individuals with $Y_{i,t}^{(0)}=0$ are \textit{survivors}.
	
	Assumptions 1 and 2 formally impose some elementary properties of
	the factual and counterfactual outcomes.
	\theoremstyle{definition} \newtheorem*{A01}{Assumption 1 (Absorbing
		State)} \begin{A01} $Y_{i,t}$ is a binary random variable and
		$Y_{i,t}=1$ implies $Y_{i,s}=1$ for all $t\leq s$. The same
		holds for the potential outcomes $Y_{i,t}^{(0)}$. \end{A01}
	
	\theoremstyle{definition} \newtheorem*{A02}{Assumption 2 (No
		Anticipation/Spill-Overs)} \begin{A02} i. For all $t\leq t^{*}$, $Y_{i,t}=Y_{i,t}^{(0)}$,
		ii. For all $t^{*}<t$, if $G_i\neq1$ then $Y_{i,t}=Y_{i,t}^{(0)}$. \end{A02}
	
	Assumption 1 states that having an outcome of $1$ is absorbing state.
	This means that if an individual has an outcome of $1$ at time $t$,
	then that individual's outcome is equal to $1$ in all future periods,
	and similarly for potential outcomes. This captures the notion that $Y_{i,t}=1$ indicates that some spell has ended. 
	
	Assumption 2 imposes conditions on potential outcomes that are standard
	in difference-in-differences. 2.i requires that
	individuals do not anticipate treatment, and so observed outcomes
	in periods weakly prior to $t^{*}$ (and thus strictly prior to the treatment time) are identical to those in the
	counterfactual world in which there is no intervention. A no-anticipation condition was introduced in causal duration analysis by \cite{Abbring2003}. 2.ii imposes that the treatment of individuals
	in Group $1$ does not impact (spill-over onto) individuals in other groups. However, we  do not rule out spill-overs between individuals in the same group.
	
	\subsection{Standard Diff-in-Diff and Related Methods}
	
	Difference-in-differences identifies causal effects under an assumption that average changes in outcomes are the same within both groups. That is, in each period $E[\Delta Y_{i,t}^{(0)}|G_i=1]=E[\Delta Y_{i,t}^{(0)}|G_i=2]$, where $\Delta$ indicates first-differencing. Equivalently, there is a constant $c$ so that for all $1\leq t\leq T$,
	 	\begin{equation}
	 	E[Y_{i,t}^{(0)}|G_i=1]=E[Y_{i,t}^{(0)}|G_i=2]+c.\label{eq:identass-2}
	 \end{equation}
	Suppose $1\leq t^{*}<T$ so there are at least two periods of data, one pre-treatment and one post-treatment. Then Assumptions 1 and 2  and the condition above identify the average
	treatment effect $\tau_{t}$. 
	
	Diff-in-diff is one example of a general class of methods that identify causal effects by assuming that there exists a fixed linear relationship between the counterfactual mean outcomes in different groups. That is, there exist parameters $\beta_1$,$\beta_2$,...,$\beta_{k}$ so that for any $1\leq t\leq T$,
	\begin{equation}
		E[Y^{(0)}_{i,t}|G_i=1]=\beta_1+\sum_{k=2}^{K}\beta_{k}E[Y^{(0)}_{i,t}|G_i=k]\label{SCstandard}.
	\end{equation}
	Imposing additional conditions on the coefficients yields alternative identification approaches. With two groups, difference-in-differences specializes the above by fixing $\beta_2=1$, in which case (\ref{eq:identass-2}) holds with $c=\beta_1$. Another case that fits into this framework is triple differences (\cite{Gruber1994}), which corresponds to $K=4$, $\beta_2=\beta_3=1$, $\beta_4=-1$, and $\beta_1$ is left unrestricted.\footnote{Imposing this, we obtain that for some fixed $\beta_1$, 	\[(E[Y_{i,t}^{(0)}|G_i=1]-E[ Y_{i,t}^{(0)}|G_i=2])-(E[ Y_{i,t}^{(0)}|G_i=3]-E[Y_{i,t}^{(0)}|G_i=4])=\beta_1,\] or equivalently
		\begin{equation*}
		E[\Delta Y_{i,t}^{(0)}|G_i=1]-E[ \Delta Y_{i,t}^{(0)}|G_i=2]=E[\Delta Y_{i,t}^{(0)}|G_i=3]-E[\Delta Y_{i,t}^{(0)}|G_i=4].
	\end{equation*}}
	Constraining $\beta_1=0$ and that the remaining coefficients are weakly positive yields a synthetic control model (\cite{Abadie2015}).

	\subsection{Consequences of the Diff-in-Diff Assumption}
	
	
	In the settings that we consider in this paper, the parallel trends assumption (\ref{eq:identass-2}) can be highly problematic. We consider cases in which the outcome is a binary indicator that an individual has reached an absorbing state by a given period. Therefore, the group mean outcome is the share of individuals in the group who have reached the absorbing state. These shares cannot exceed $1$ and are increasing with time. As such, they tend to converge, regardless of treatment. That is, the magnitude of the difference in group-mean outcomes tends to decrease, even if treatment has no effect. Standard diff-in-diff would erroneously interpret such a decrease as evidence of a treatment effect, leading to biased and inconsistent causal estimates.
	
	
	More precisely, consider that parallel trends (a) places a ceiling on the shares in each group that can ever enter the absorbing state, and (b) implies the rates at which survivors reach the absorbing state diverges between groups. To elaborate, consider the unemployment example and suppose that prior to treatment, $80\%$ are employed in Group $1$ and $60\%$ in Group $2$. To see (a) in action, note that by parallel trends and the fact that percentages cannot exceed $100$, no more that $80\%$ of Group $2$ can ever exit unemployment. Formally, by (\ref{eq:identass-2}) the counterfactual share that reaches the absorbing state cannot exceed $1+c$ in Group 1 and $1-c$ in Group $2$. To explain (b), note that the pre-treatment share unemployed in Group $2$ is $40\%$, which is twice that in Group $1$. For the difference to remain constant into the next period, individuals in Group $1$ who are unemployed at time $t$ must be twice as likely to gain a job by $t+1$ as those in Group $2$. In a later period, if $90\%$ in Group $1$ are employed and $70\%$ in Group $2$, individuals in group $1$ must find jobs at three times the rate in Group $2$. We formalize this point in Section A.1 of the supplementary appendix.

	Figure 1.1(a) provides a graphical illustration. The figure shows population mean factual and counterfactual outcomes under a duration model specified in Section C of the supplementary appendix. Factual mean outcomes, i.e., shares of the population that have reached the absorbing state, are plotted over time by solid lines, blue for Group $1$ and red for Group $2$. The mean outcome for Group $1$ under a counterfactual of no treatment, is plotted by the dashed blue line. 
	
	\begin{figure}[H]
		\caption{Deviations from Parallel Trends}
		
		\subfloat[Mean Outcomes]{
			
			\includegraphics[scale=0.23]{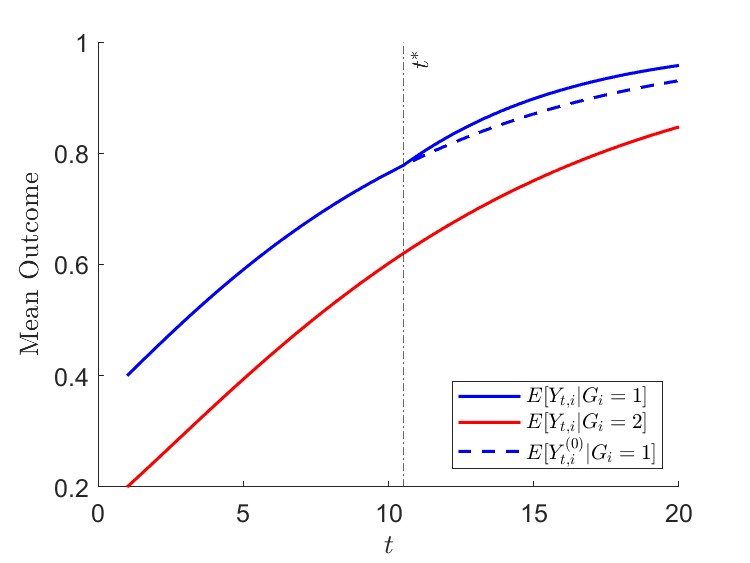}}\subfloat[Transformed Mean Outcomes]{\includegraphics[scale=0.23]{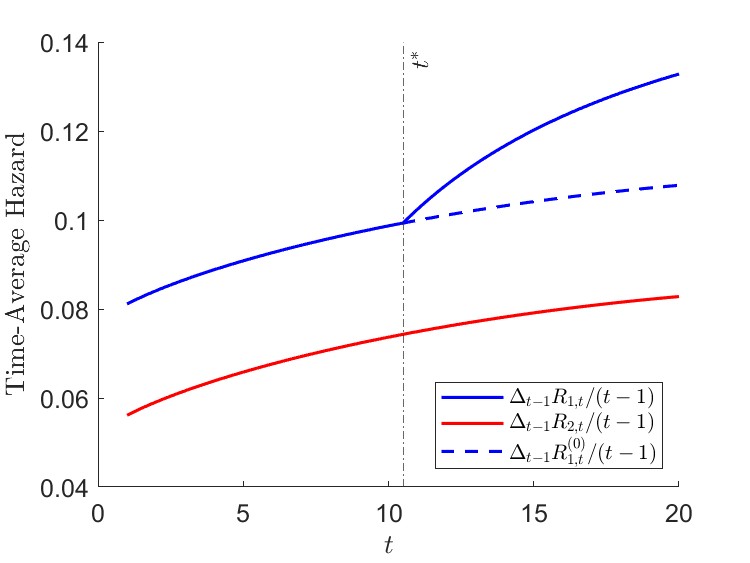}}
	\end{figure}
	
	In the figure, we see that mean outcomes for Group $1$ are greater than for Group $2$ in the initial period. Due to the duration nature of the setting, the difference decreases over time under the counterfactual of no intervention. In this example, the convergence is of a sufficient magnitude that the average difference in the observable factual outcomes is smaller over the post-treatment period than in the pre-treatment period. Therefore, in expectation standard diff-in-diff will estimate a negative treatment effect. However, the true treatment effect is positive, as seen by the positive gap between the solid and dashed blue lines. In fact, in this model the bias of standard diff-in-diff is roughly four times the value of the true treatment effect.

	\section{Model and Identification}
	
	The discussion in the previous section suggests that the assumptions we use to identify and estimate causal effects should be compatible with the type of convergence shown in Figure 1.1(a). As an alternative, we assume parallel trends or some other fixed linear relationship holds between group-specific hazard rates rather than mean outcomes. Parallel trends in the hazard rates is consistent with the convergence of counterfactual mean outcomes. Indeed, the assumption holds in Figure 1.1(a) and thus the convergence in the figure does not result in misleading inference under this assumption.
	
	A hazard rate is a rate at which individuals who have not yet reached the absorbing state, reach that state. Let $\delta>0$ and consider the probability that, under the counterfactual of no intervention, an individual in Group $k$ reaches the absorbing state between times
$t$ and $t+\delta$, conditional on having not having reached it prior to $t$.  If we scale this probability by $1/\delta$, then the limit as the increment $\delta$ shrinks to zero is the counterfactual hazard rate for Group $k$. Formally, the counterfactual hazard for Group $k$ at time $t$ is
\[
h^{(0)}_{k}(t):=\lim_{\delta\downarrow0}\frac{P(Y_{i,t+\delta}^{(0)}=1|Y_{i,t}^{(0)}=0,G_i=k)}{\delta}.
\]
 We can define the factual hazard rate $h_k(t)$ analogously by replacing counterfactual outcomes with factual outcomes in the definition.
	
	 Throughout, we continue to take $k=1$ to be the unique treated group (for an extension to staggered adoption see Section A.3 of the supplementary appendix). In the two-group case, we consider two leading specifications. The first is the \emph{common dynamics} assumption. This states that, between times $1$ and $T$, changes in the counterfactual hazard rate over any time interval are equal across groups:
	\begin{equation}
		h^{(0)}_{1}(t)-h^{(0)}_{1}(s)= h^{(0)}_{2}(t)-h^{(0)}_{2}(s), \label{eq:hlevel0}
	\end{equation}
	for all $1\leq s\leq t\leq T$. The condition allows for common shocks that impact the counterfactual hazards of individuals in both the treated and untreated groups, so long as these shocks induce parallel movements in the hazard rates. It also allows for the possibility that the hazard rates evolve over time, increasing or decreasing with the time spent in the $Y_{i,t}=0$ state.\footnote{The condition is only required to hold between periods $1$ and $T$ rather than indefinitely far into the past and or future and thus leaving initial survivals $E[1-Y_{i,1}|G_i=k]$ unrestricted.}
	
	The second is a \emph{proportional hazard} assumption, as implied by the models of \cite{Hunt1995} and \cite{Wu2022}. This states that counterfactual hazards are strictly positive and proportional changes over any time interval are equal across groups
	\begin{equation}
		h^{(0)}_{1}(t)/h^{(0)}_{1}(s)= h^{(0)}_{2}(t)/h^{(0)}_{2}(s),\label{prophaz0}
	\end{equation}
	for all $1\leq s\leq t\leq T$. This restriction means that shocks and time trends leave the difference in log counterfactual hazard rates unchanged.

	Both the common dynamics condition (\ref{eq:hlevel0}) and proportional hazards condition (\ref{prophaz0}) imply that a fixed linear relationship holds between the counterfactual hazard rates of the two groups. Condition (\ref{eq:hlevel0}) is equivalent to the existence of a fixed constant $c$ such that 
	\begin{equation}h^{(0)}_{1}(t)=h^{(0)}_{2}(t)+c,\,\,\,\forall 1\leq t\leq T \label{eq:hlevel}
	\end{equation}
	and similarly, (\ref{prophaz0}) is equivalent to the existence of some $c$ so that
		\begin{equation}h^{(0)}_{1}(t)=c h^{(0)}_{2}(t),\,\,\,\forall 1\leq t\leq T. \label{prophaz}
	\end{equation}
	
	Note that if counterfactual hazard rates are identical across groups, that is  $h^{(0)}_{1}(t)=h^{(0)}_{2}(t)$, then both the common dynamics and proportional hazards conditions hold. Even in this case, mean outcomes may differ across groups if the initial proportions of individuals in the absorbing state differ.

	We also allow for more general linear restrictions between counterfactual hazard rates which may be more appropriate in settings with one treated group and many control groups. Consider assumptions of the form
	\begin{equation}
		h^{(0)}_{1}(t)=\beta_1+\sum_{k=2}^{K}\beta_{k}h^{(0)}_{k}(t) ,\,\,\,\forall1\leq t\leq T.\label{SCeq1}
	\end{equation}
	where the parameters $\beta_1,...,\beta_{k}$ are constant over time. The common dynamics assumption (\ref{eq:hlevel}) corresponds to $K=2$ with $\beta_2=1$ and $\beta_1$ relabelled as $c$. The proportional hazard restriction (\ref{prophaz}) corresponds to $K=2$ and $\beta_1=0$ and $\beta_2$ renamed as $c$. The general form also allows for a triple differences-type restriction on the hazard rates, that is 
	\[
	[h^{(0)}_{1}(t)-h^{(0)}_{2}(t)]-[h^{(0)}_{3}(t)-h^{(0)}_{4}(t)]=c,
	\]
	which corresponds to (\ref{SCeq1}) with $\beta_2=\beta_3=1$, $\beta_4=-1$, and $\beta_1$ relabelled as $c$.  Moreover, restricting $\beta_1=0$ and the remaining coefficients to be weakly positive and sum to $1$, one obtains a condition on hazard rates analogous to synthetic control (\cite{Abadie2015}).
	
	We show that linear restrictions on the hazard rates allow for simple and transparent identification and estimation methods that  retain desirable features of  the corresponding diff-in-diff and related methods. The key observation is that restrictions of the form (\ref{SCeq1}) on the hazard rates translate into the same linear restrictions on objects that can be written as known transformations of mean outcomes, and are therefore directly estimable from the data.
	
	
	\subsection{Identification}
	
	We establish identification of causal effects under linear restrictions on group-specific hazard rates of the form in (\ref{SCeq1}). This includes the special cases of common dynamics (\ref{eq:hlevel}) and the proportional hazard restriction (\ref{prophaz}).
	 We first connect the restrictions on the hazard rates to restrictions on observable quantities.  Define the counterfactual \emph{negative log survival} by
	\begin{equation}
		R^{(0)}_{k,t}:=-ln\big(1-E[Y_{i,t}^{(0)}|G_i=k]\big). \label{eq:Rdef}
	\end{equation}
	The factual counterpart $R_{k,t}$ is defined analogously by replacing counterfactual outcomes with observed outcomes, that is
	\begin{equation}
		R_{k,t}:=-ln\big(1-E[Y_{i,t}|G_i=k]\big). \label{eq:Rfactdef}
	\end{equation}
	It is a standard result in duration analysis that the integrated hazard is equal to negative log survival. Applying this result here, we have that under Assumption 1 for any $t$ with $1< t\leq T$
	\begin{equation}
		\Delta R^{(0)}_{k,t}:=R^{(0)}_{k,t}-R^{(0)}_{k,t-1}=\int_{t-1}^{t}h^{(0)}_{k}(r)dr. \label{eq:Rint}
	\end{equation}
	That is, the first difference $\Delta R^{(0)}_{k,t}$ is  the integrated hazard rate over the time from $t-1$ to $t$. Linear restrictions on counterfactual hazard rates thus imply linear restrictions on changes in $R^{(0)}_{k,t}$. We can therefore restate our identifying assumptions in terms of $R^{(0)}_{k,t}$. Under Assumption 1 and common dynamics (\ref{eq:hlevel}) we have 
	\begin{equation}
		\Delta R^{(0)}_{1,t}-\Delta R^{(0)}_{1,t-1}=\Delta R^{(0)}_{2,t}-\Delta R^{(0)}_{2,t-1}\label{CD1}
		\end{equation}
	 for any whole number $t$ so that $1<t\leq T$. That is, one period \textit{changes} in $R^{(0)}_{k,t}$ satisfy a parallel trends condition across groups. 	In the special case in which the hazard rates are equal across groups $c=0$, we obtain $\Delta R^{(0)}_{1,t}=\Delta R^{(0)}_{2,t}$, so that \textit{levels} of $R^{(0)}_{k,t}$  exhibit parallel trends.
	
Under Assumption 1 and the proportional hazard assumption (\ref{prophaz}), changes in $R^{(0)}_{k,t}$ across the two groups are proportional across groups. That is, 
\begin{equation}
	\Delta R^{(0)}_{1,t}/\Delta R^{(0)}_{1,t-1}=\Delta R^{(0)}_{2,t}/\Delta R^{(0)}_{2,t-1}. \label{PH1}
\end{equation}
Again, if counterfactual hazards are identical across groups, then $c=1$ and we again obtain parallel trends in levels of $R^{(0)}_{k,t}$.

Equations (\ref{CD1}) and (\ref{PH1}) each suggest a simple identification argument. Under (\ref{CD1}), one can apply difference in differences to $\Delta R_{k,t}$ (or its log under (\ref{PH1})) in order to impute post-treatment values of $\Delta R^{(0)}_{1,t}$ and thus of  $R^{(0)}_{1,t}$. The logarithmic function in (\ref{eq:Rdef}) can then be inverted to recover $E[Y_{i,t}^{(0)}|G_i=1]$ and the ATT.

In order to apply difference in differences to \textbf{changes} in $R_{k,t}$ (or its logarithm), we must be able to calculate $\Delta R_{1,t}$ for some pre-treatment period $t$. Thus we require at least two periods of pre-treatment data and hence at least three periods of data in all. This deviates from standard difference in differences which requires only two periods of data. An exception is the case of equal hazard rates ($c=0$ in (\ref{eq:hlevel}) or $c=1$ in (\ref{prophaz})), in which case levels of $R^{(0)}_{k,t}$ exhibit parallel trends.


As a worked example, consider the case in which there are three periods of data $t=1,2,3$ with treatment occurring between the final two periods. The key step is to identify $ R^{(0)}_{1,3}$. Inverting (\ref{eq:Rdef}) one obtains
\[E[Y^{(0)}_{i,3}|G_i=1]=1-exp(- R^{(0)}_{1,3}).\]
The ATT is then obtained by subtracting the above from the factual mean outcome $E[Y_{i,3}|G_i=1]$.

To establish identification of $R^{(0)}_{1,3}$, note that under Assumptions 1 and 2, $ R^{(0)}_{1,t}=R_{1,t}$ for $t=1,2$, and by Assumption 2.ii, $R^{(0)}_{2,t}=R_{2,t}$ for $t=1,2,3$. So by (\ref{CD1}),
\[R^{(0)}_{1,3}=R_{1,2}+\Delta R_{1,2}+\Delta R_{2,3}-\Delta R_{2,2},\]
or alternatively, from (\ref{PH1}),
\[R^{(0)}_{1,3}=R_{1,2}+\Delta R_{1,2}\Delta R_{2,3}/\Delta R_{2,2}.\]

	\subsubsection{Identification Under Common Dynamics/Proportional Hazards}

In order to present results that motivate estimators for the $T>3$ case, it is convenient to express (\ref{CD1}) equivalently as follows. Define $\Delta_s R^{(0)}_{k,t}:=R^{(0)}_{k,t}-R^{(0)}_{k,t-s}$ and similarly $\Delta_s R_{k,t}:=R_{k,t}-R_{k,t-s}$. There is a constant $c$ so that for each $1< t\leq T$
\begin{equation}
	\frac{\Delta_{t-1} R^{(0)}_{1,t}}{t-1}=c+\frac{\Delta_{t-1} R^{(0)}_{2,t}}{t-1}. \label{eq:RlevelDiff2}
\end{equation}
In fact, Assumption 1 and common dynamics imply that the equation above holds not only for  whole numbers but for any real number $t\in(1,T]$. Thus this restriction is also applicable to settings with unequal time increments (see Remark 3 later in this section). We refer to $\Delta_{t-1} R^{(0)}_{1,t}/(t-1)$ as the `time-average hazard'.
The constant $c$ in (\ref{eq:RlevelDiff2}) is identical to that in (\ref{eq:hlevel}).
Similarly, it is convenient to express (\ref{PH1}) as,
\begin{equation}
		{\Delta_{t-1} R_{1,t}}=c {\Delta_{t-1} R_{2,t}} \label{propeq}
\end{equation}
where the constant $c$ is identical to that in the condition (\ref{prophaz}). Under Assumption 1 and proportional hazards, the above holds for any $t\in(1,T]$.

	\theoremstyle{plain} \newtheorem*{2g}{Theorem 1} \begin{2g}
	Suppose Assumptions 1 and 2 hold and there is data from at least two pre-treatment periods. Under the common dynamics condition (\ref{eq:hlevel}), $c$ and $R^{(0)}_{1,t}$ are identified by,
	\begin{align}
	c&=\frac{\Delta_{t-1} R_{1,t}}{t-1}-\frac{\Delta_{t-1} R_{2,t}}{t-1},\,\,\forall 1<t \leq t^* \label{id1}\\
  R^{(0)}_{1,t}&=R_{1,1}+\Delta_{t-1} R_{2,t}+(t-1)c,\,\,\forall t \leq T.
\end{align} 
If proportional hazards (\ref{prophaz}) holds instead of (\ref{eq:hlevel}) and if  $E[Y_{i,t}|G_i=2]\neq E[Y_{i,1}|G_i=2]$ for some $t\leq t^*$,\footnote{This ensures that $\Delta_{t-1} R_{2,t}\neq0$ for at least one value of $1<t\leq t^*$ and so $\Delta_{t-1} R_{1,t}=c\Delta_{t-1} R_{2,t}$ can be inverted to get $c=\Delta_{t-1} R_{1,t}/\Delta_{t-1} R_{2,t}$.} then $c$ and $R^{(0)}_{1,t}$ are identified by
	\begin{align}
{\Delta_{t-1} R_{1,t}}&=c{\Delta_{t-1} R_{2,t}},\,\,\forall 1<t \leq  t^*\label{idPH}\\	 R^{(0)}_{1,t}&=R_{1,1}+c\Delta_{t-1} R_{2,t},\,\,\forall t \leq T
\end{align}
	In either case, the counterfactual mean outcome is then identified by
		\begin{equation}
	E[Y^{(0)}_{i,t}|G_i=1]=1-exp(-R^{(0)}_{1,t}). \label{finalchar}
\end{equation}
\end{2g}

For a graphical illustration of the identification strategy in Theorem 1, consider Figures 1.1(a) and 1.1(b). These figures correspond to a continuous time duration model in which the common dynamics condition holds and we use this same model as the data-generating process in the Monte Carlo exercises in Section C of the supplementary appendix. The solid curves in Figure 1.1(a) plot the conditional mean outcomes over time for the two groups, and in 1.1(b) they plot the time-average hazard $\Delta_{t-1} R_{k,t}/(t-1)$. Note that in both diagrams, the solid curves represent factual (rather than counterfactual) objects, and are thus directly estimable. 

Parallel trends clearly fails in 1.1(a), but it holds for the curves in 1.1(b). This follows by (\ref{eq:RlevelDiff2}) which implies that in the pre-treatment period, the gap between the solid curves in 1.1(b) is constant and equal to $c$. The post-treatment $\Delta_{t-1}R^{(0)}_{1,t}/(t-1)$, which is indicated by the dashed blue curve in 1.1(b) can then be imputed by extrapolating this constant gap to the post-treatment period. With these counterfactual differences in hand, one can invert the mapping from the curves in 1.1(a) to 1.1(b) to obtain $E[Y_{i,t}^{(0)}|G_i=1]$ which is plotted by the dashed blue curve in 1.1(a). The difference between this curve and the solid red curve in 1.1(a) gives the ATT in each period.

\newtheorem*{remark1}{Remark 1: Over-Identification}
\begin{remark1}
 Note that if $t^*>2$ then causal effects are over-identified. In particular, (\ref{id1}) and (\ref{idPH}) each constitute $t^{*}-1$ equations from which to identify $c$. It is this over-identification that we exploit to test pre-treatment parallel trends, analogously with standard difference in differences.
\end{remark1}

\newtheorem*{remark2}{Remark 2: Equivalent Characterizations}
\begin{remark2}
Theorem 1 characterizes $c$ in terms of scaled long differences in negative log survivals. However, one can combine differences to obtain equivalent characterizations of $c$ and  $E[Y^{(0)}_{i,t}|G_i=1]$. For example, the equations (\ref{id1}) are equivalent to either of the following two systems of equations
\begin{align*}
	c&=\Delta R_{1,t}-\Delta R_{2,t},\,\,\forall 1<t \leq t^*\\
	c&=(\Delta_{t} R_{1,t^{*}}-\Delta_{t} R_{2,t^{*}})/t,\,\,\forall 1\leq t \leq t^*-1.
\end{align*}
The characterization in the Theorem has the advantage that (a) it generalizes straight-forwardly to cases in which $t$ is observed at uneven increments (see Remark 3 below), and (b) estimates of $\Delta_{t-1} R_{k,t}/(t-1)$ are typically less noisy than those of $\Delta R_{k,t}$ (the former is the average of $\Delta R_{k,s}$ over $s=2,...,t$), which allows for easier visual inspection of pre-treatment  parallel/proportional trends. 
\end{remark2}

\newtheorem*{remark3}{Remark 3: Uneven Increments and Repeated Cross-Sections}
\begin{remark3}
	The results in Theorem 1 can be straight-forwardly adapted to cases in which observations are made at uneven increments. That is, observations are made at times $t\in\mathcal{T}$ where $\mathcal{T}$ is some finite set of real numbers whose smallest element is $1$, rather than $t\in\{1,2,...,T\}$. In this case, the equations that identify $c$ in (\ref{id1}) and (\ref{idPH}) should be understood to hold for each $t\in\mathcal{T}\setminus 1$.
	
	The identification results in Theorem 1 depend on the distribution of observables only on group-specific mean outcomes $E[Y_{i,t}|G_i=k]$. These objects are identified and estimable with data on repeated cross sections rather than panel data.
\end{remark3}

\newtheorem*{remark4}{Remark 4: Comparison to \cite{Wooldridge2023}}

\begin{remark4}
The conditions (\ref{eq:RlevelDiff2}) and (\ref{propeq}) closely relate to specifications in \cite{Wooldridge2023}. To connect our results to the analysis in that paper, first note that equations (\ref{eq:RlevelDiff2}) imply that there exist group-specific constants $\xi_k$ and $c_k$ (with $c_1$ normalized to $0$), and time fixed effects $\gamma_t$ so that for $t=1,...,T$
\begin{equation}
	R^{(0)}_{k,t}=\xi_{k}-c_{k}\cdot t+\gamma_t. \label{eq:Rlevel}
\end{equation}
This is a two-way fixed effects model with group-specific linear trends in $R^{(0)}_{k,t}$. Similarly, equation (\ref{propeq}) implies there are $\xi_k$, $c_k$ (with $c_1$ normalized to $1$), and $\gamma_t$ so that
\begin{equation}
	R^{(0)}_{k,t}=\xi_{k}+c_{k}\cdot \gamma_t. \label{eq:Rprop1}
\end{equation}
If $c_2=0$ in (\ref{eq:Rlevel}) or $c_2=1$ in (\ref{eq:Rprop1}), then we obtain
\begin{equation}
	R^{(0)}_{k,t}=\xi_{k}+\gamma_t \label{eq:Requalhaz}
\end{equation}
which is a standard two-way fixed effects specification for $R^{(0)}_{k,t}$. Note that in the duration setting, $c_2=0$ in (\ref{eq:Rlevel}) and $c_2=1$ in (\ref{eq:Rprop1}) correspond to the case in which hazard rates are equal across groups. From the definition of $R^{(0)}_{k,t}$, one sees that (\ref{eq:Requalhaz}) is equivalent to \[1-E[Y_{i,t}^{(0)}|G_i=k]=\exp(-\xi_{k}-\gamma_t).\]
This is precisely the exponential model given by equations (2.11) and (2.12) in  \cite{Wooldridge2023} (albeit with $1-Y_{i,t}$ in place of $Y_{i,t}$). More generally, when $c_2\neq0$, (\ref{eq:Rlevel}) can be obtained by including heterogeneous linear time trends to the exponential model in Wooldridge, and when $c_2\neq1$, (\ref{eq:Rprop1}) can be obtained by including multiplicative group-time effects. Section 4.2 in \cite{Wooldridge2023} suggests inclusion of heterogeneous linear time trends as a correction for violation of parallel trends.\footnote{We thank an anonymous referee for pointing out these connections.}

While our analysis closely relates to \cite{Wooldridge2023}, we arrive at (\ref{eq:RlevelDiff2}) and (\ref{propeq}) from a distinct starting point. In \cite{Wooldridge2023}, the exponential model for mean outcomes is motivated by a latent variable framework for limited dependent variables, and the link function is determined by a distributional assumption on unobserved noise. In our setting, the exponential transformation in the definition of $R^{(0)}_{k,t}$ arises from the duration structure of the data: the fact that $R^{(0)}_{k,t}$ equals an integral of the counterfactual hazard follows from the absorbing state property, not from a distributional assumption. This provides a structural foundation for the exponential models above in settings with duration outcomes.

\end{remark4}
	
\subsubsection{Identification Under General Linear Restrictions}	
	
Recall that both the common dynamics and proportional hazards assumptions specialize the general linear restriction (\ref{SCeq1}). The general linear restriction implies an analogous linear relationship, between time-average hazards
\[
\frac{\Delta_{t-1} R^{(0)}_{1,t}}{t-1}=\beta_1+\sum_{k=2}^{K}\beta_{k}\frac{\Delta_{t-1} R^{(0)}_{k,t}}{t-1},\,\,\,\,\forall 1<t\leq T.
\]	When $K=2$ this nests the common dynamics and proportional hazards restrictions, the former with $\beta_1=c$ and $\beta_2=1$, the latter with  $\beta_1=0$ and $\beta_2=c$.

Theorem 1 is a special case of Theorem 2 below, which considers the case of multiple untreated groups and general linear restrictions in (\ref{SCeq1}). The result can accommodate  additional a priori restrictions on the coefficients. For example, in the case of two groups, the restriction $\beta_2=1$ recovers the common dynamics condition. We capture such additional restrictions by requiring that the vector of coefficients $(\beta_1,\beta_2,...,\beta_K)$ belongs to a set $\mathcal{B}$. Causal effects may be identified without imposing any such restrictions if there are sufficiently many pre-treatment periods.
		\theoremstyle{plain} \newtheorem*{2f}{Theorem 2} \begin{2f}
		Suppose Assumptions 1 and 2 hold. In addition suppose that (\ref{SCeq1}) holds with $(\beta_1,\beta_2,...,\beta_K)\in\mathcal{B}$ for a known set $\mathcal{B}$.  Then $\beta_1,\beta_2,...,\beta_K$ satisfy the equations
		\begin{equation}
\frac{\Delta_{t-1} R_{1,t}}{t-1}=\beta_1+\sum_{k=2}^{K}\beta_{k}\frac{\Delta_{t-1} R_{k,t}}{t-1},\,\,\forall  1<t\leq t^{*}. \label{coeffid}
		\end{equation}
If the solution to the equations is unique within $\mathcal{B}$ (i.e., the equations identify $\beta_1,\beta_2,...,\beta_K$), then for $1<t\leq T$, $\Delta R^{(0)}_{1,t}$ is identified by 
$
	R^{(0)}_{1,t}=R_{1,1}+(t-1)\beta_1+\sum_{k=2}^{K}\beta_{k}\Delta_{t-1}R_{k,t}
$ and the counterfactual mean outcome by (\ref{finalchar}).
	\end{2f}
	
Theorem 2 identifies causal effects only if there is a unique solution to the equations (\ref{coeffid}) that satisfy the constraints incorporated in $\mathcal{B}$. If we do not constrain the coefficients $\{\beta_{k}\}_{k=1}^K$, then a necessary condition for uniqueness of the solution to (\ref{coeffid}), and thus for identification, is that $t^*>K$ and uniqueness is generic whenever this holds. Given a sufficient number of pre-treatment periods, the parameters in (\ref{coeffid}) are over-identified. This suggests we can test the identifying restrictions using say, placebo tests.

\subsection{Incorporating Covariates}

In difference-in-differences analysis, covariate adjustment presents a means of credibly recovering causal effects when parallel trends fails. If the distribution of covariates differs between groups, then parallel trends may fail in aggregate, even if it holds within each covariate stratum. This motivates the covariate balancing method of \cite{Abadie2005} in standard (i.e., non-duration) difference-in-differences. In this section we establish identification results that employ covariate adjustment to correct for failure of the common dynamics, proportional hazards, or other linear restrictions in duration settings. We first discuss the case in which the relevant assumptions do not hold in aggregate but hold within each covariate stratum. We then present results that apply only under additional assumptions but which motivate simpler estimation methods.

Let $X_i$ be a vector of time-invariant individual characteristics. Suppose that the common dynamics, proportional hazards, or other restrictions hold within each stratum of $X_i$. To state this formally, let  $h_{k}^{(0)}(t;x)$ denote the Group $k$  counterfactual hazard rate at time $t$ for the sub-population for whom $X_{i}$ is equal to $x$, that is
\[
h_{k}^{(0)}(t;x)=\lim_{\delta\downarrow0}\frac{P(Y_{i,t+\delta}^{(0)}=1|Y_{i,t}^{(0)}=0,G_i=k,X_{i}=x)}{\delta}.
\]
In the case of two groups, covariate-conditional versions of the common dynamics and proportional hazard conditions are given below by (\ref{covCD}) and (\ref{covPH}) respectively.
\begin{align}
	h_{1}^{(0)}(t;x)=h_{2}^{(0)}(t;x)+c(x)\label{covCD}\\
	h_{1}^{(0)}(t;x)=c(x)h_{2}^{(0)}(t;x)\label{covPH}.
\end{align}
for a time-invariant function $c(\cdot)$.  More generally, one may assume a time-invariant linear relationship between covariate-conditional hazards, that is
\begin{equation}
	h_{1}^{(0)}(t,x)=\beta_{1}(x)+\sum_{k=2}^{K}\beta_{k}(x)h_{k}^{(0)}(t,x).\label{generalCOV}
\end{equation}
Much as in the case without covariates, with $K=2$ the above yields (\ref{covCD}) and (\ref{covPH}) as special cases, in particular $\beta_2(x)=1$ or $\beta_1(x)=0$ respectively.
 
In order to impute conditional mean counterfactual outcomes for the treated group, one can apply the identification results in Theorems 1 and 2 separately within each covariate stratum. To be precise, define the \textit{conditional negative log survival} by 
 \begin{align*}
 	R_{k,t}^{(0)}(x)&:=-ln\big(1-E[Y_{i,t}^{(0)}|G_i=k,X_i=x]\big)\\
 	R_{k,t}(x)&:=-ln\big(1-E[Y_{i,t}|G_i=k,X_i=x]\big).
 \end{align*}  
 Then one can impute $R_{1,t}^{(0)}(x)$ for a post-treatment period $t$ much as one imputes $R_{1,t}^{(0)}$ in Theorems 1 and 2. The distinction is that the conditional negative log survivals replace their unconditional counterparts, and the conditional coefficient $c(x)$ (or coefficients $\beta_1(x),\beta_2(x),...,\beta_K(x)$) replace the coefficient $c$ (or $\beta_1,\beta_2,...,\beta_K$). Counterfactual conditional mean outcomes are then given by 
 \begin{equation*}
 	E[Y^{(0)}_{i,t}|G_i=1,X_i=x]=1-exp(- R^{(0)}_{1,t}(x)).
 \end{equation*}
 A formal theorem stating these results is located in Section A.2 of the supplementary Appendix. This identification result suggests a plug-in non-parametric estimator. In particular, one may replace the conditional means of the form $E[Y_{it}|G_i=k,X_i=x]$ in the conditional negative log survival, with non-parametric regression estimates. This may be infeasible if sample sizes are insufficiently large. This motivates approaches that are valid under additional or alternative assumptions. We propose a covariate balancing approach that is valid under further restrictions on the relationship between covariate-conditional hazard rates but leaves the dependence of group-specific conditional hazards on the covariates unrestricted. Additionally, in Appendix A.4 we propose a semiparametric approach that assumes the covariates enter the conditional hazard function multiplicatively via a linear index and link function.


\subsubsection{Covariate Balancing Weights}

We specify a non-parametric approach that may repair a failure of parallel trends (or other linear restrictions) by balancing the distributions of the covariates between groups, similar to \cite{Abadie2005}. More precisely, we balance the distribution of covariates among those who have not yet reached the absorbing state in the initial period. Define a weight function $\omega_{k}$ on the support of the covariates. In the case of continuous covariates one can replace the conditional
probabilities with conditional probability densities and obtain a weighting of the kind developed in \cite{DiNardo1996}.
\begin{equation}
	\omega_{k}(x):=\frac{P(X_{i}=x|Y_{i,1}=0,G_i=1)}{P(X_{i}=x|Y_{i,1}=0,G_i=k)}.\label{weight1}
\end{equation}
The weighting function can be written equivalently in terms of a type of  propensity score (\cite{Rosenbaum1983}). Let $p_k(x)$ be the probability that an individual is in Group $k$ given they have covariate values $x$ and have not reached the absorbing state at the initial time. Formally, we define
\[p_{k}(x):=P(G_{i}=k|Y_{i,1}=0,X_{i}=x).\]
Then applying Bayes' rule we obtain the following which is only well-defined if $p_k(x)>0$, 
\begin{equation}
	\omega_{k}(x)=\frac{p_{1}(x)P(G_{i}=k|Y_{i,1}=0)}{p_{k}(x)P(G_{i}=1|Y_{i,1}=0)}.\label{weight2}
\end{equation}
By weighting individuals in the untreated group by $\omega_{k}(X_{i})$ we down-weight those individuals with values of the covariates that are more prevalent among Group $k$ than Group $1$, and up-weight those whose values are less prevalent.

The counterfactual hazard function for the weighted group-$k$ population is
\begin{equation}\tilde{h}_{k}(t):=\lim_{\delta\downarrow0}\frac{E[\omega_{k}(X_{i})Y_{i,t+\delta}|G_i=k]-E[\omega_{k}(X_{i})Y_{i,t}|G_i=k]}{\delta  E[\omega_{k}(X_{i})(1-Y_{i,t})|G_i=k]}.\label{htildedef}\end{equation}
This \textit{weighted hazard function} is the hazard function for a hypothetical population with the same covariate-conditional hazards as Group $k$, in which the distribution of covariates (among those not in the absorbing state at time $1$) is identical to that of Group $1$. Thus any differences between $\tilde{h}_{k}(t)$ and $\tilde{h}_{1}(t)$ cannot be attributed to difference in the distribution of covariates and must be instead due to group-level factors or differences in the distribution of unobservables. Indeed, if $\tilde{h}_{k}(t,x)$ is identical to  $\tilde{h}_{1}(t,x)$ for all $t$ and $x$, then the weighted hazard functions for the two groups are identical. 
The weighted counterfactual hazard $\tilde{h}_{k}^{(0)}(t)$ is defined analogously by replacing factual outcomes with counterfactual outcomes in (\ref{htildedef}).

Our re-weighting approach is valid when there is a fixed linear relationship between the weighted hazard functions of the different populations. The corresponding common dynamics and proportional hazards assumptions are respectively
\begin{equation}\tilde{h}_{1}^{(0)}(t)=\tilde{h}_{2}^{(0)}(t)+c\, \text{ and }\,\tilde{h}_{1}^{(0)}(t)=c\tilde{h}_{2}^{(0)}(t),\,\,\,\forall 1\leq t\leq T \label{weighted1}
	\end{equation}
and the more general linear restrictions by
\begin{equation}\tilde{h}_{1}^{(0)}(t)=\beta_1 + \sum_{k=2}^K \beta_k\tilde{h}_{k}^{(0)}(t),\,\,\,\forall 1\leq t\leq T.\label{weighted2}
	\end{equation}
These restrictions may be justified in two ways. Firstly, compared to the unbalanced common dynamics/proportional hazards restrictions, it may be more plausible to assume a priori that these restrictions hold only after balancing the distribution of covariates between groups, and likewise for more general linear restrictions. Thus one may hypothesise that one of the conditions in (\ref{weighted1}) or (\ref{weighted2}) holds without reference to primitive conditions and then attempt to falsify the assumption using specification tests. Secondly, as we show in Proposition 1 below, the weighted common dynamics assumption in (\ref{weighted1})  follows from a special case of the conditional common dynamics assumption (\ref{covCD}). In particular, it follows when $c(x)$ does not depend on $x$.

A sufficient condition for  $c(x)$ not to depend on $x$ is that conditional hazards are equal, i.e., $h_1(t,x)=h_2(t,x)$ for all $t$ and $x$. This holds under unconfoundedness/conditional ignorability, that is $Y_{i,t}^{(0)}\indep G_i|X_i,Y_{i,1}=0$. In words, it suffices that, among individuals in a particular covariate stratum (and who are not initially in the absorbing state), the assignment to the treatment/control group is random (i.e., unrelated to potential outcomes). Another sufficient condition is that (\ref{covCD}) holds and that covariate-specific hazard rates obey an additively separable structure. In particular, $h_k(t,x)=\bar{h}_k(t)+q(t,x)$, for $k=1,2$, where $\bar{h}_k(t)$ is a baseline hazard rate, $q(t,x)$ captures an additive impact of covariates on the hazard rate. This is an additive hazard model of the kind suggested by \cite{Aalen1989}.

\theoremstyle{plain} \newtheorem*{Prop}{Proposition 1} \begin{Prop}
	Suppose Assumption 1 holds and (\ref{covCD}) holds with $c(x)$ equal to a constant $c$. Then $\tilde{h}_{1}^{(0)}(t)=\tilde{h}_{2}^{(0)}(t)+c$.
\end{Prop}

Identification of causal effects under conditions in (\ref{weighted1}) or (\ref{weighted2}) follows similarly to the unweighted case. In particular, define the weighted negative log survival by
\[\tilde{R}_{k,t}:=-ln\big(E[\omega_k(X_i)(1-Y_{i,t})|G_i=k]\big)\]
and its counterfactual counterpart $\tilde{R}_{k,t}^{(0)}$ as above but with $Y_{i,t}^{(0)}$ replacing $Y_{i,t}$. Because $ \tilde{R}_{1,t}^{(0)}= {R}_{1,t}^{(0)}$, causal effects can then be computed just as in Theorems 1 and 2.
	\theoremstyle{plain} \newtheorem*{P6}{Theorem 3} \begin{P6}
	Suppose Assumptions 1 and 2 hold and that $p_k(x)>0$ for all $k=1,...,K$ and $x$ in the support of $X$. Suppose (\ref{weighted2}) holds with $(\beta_1,\beta_2,...,\beta_K)\in\mathcal{B}$ for a known set $\mathcal{B}$. Then  $\beta_1,\beta_2,...,\beta_K$ satisfy the equations
	\begin{equation}
		\frac{\Delta_{t-1} R_{1,t}}{t-1}=\beta_1+\sum_{k=2}^{K}\beta_{k}\frac{\Delta_{t-1} \tilde{R}_{k,t}}{t-1},\,\,\forall  1<t \leq  t^{*}. \label{coeffid334}
	\end{equation}
	If the solution to the equations is unique within $\mathcal{B}$, then ${R}^{(0)}_{1,t}$ is identified (for $1<t\leq T$) by 
	${R}^{(0)}_{1,t}=R_{1,1} +(t-1)\beta_1+\sum_{k=2}^{K}\beta_{k}\Delta_{t-1} \tilde{R}_{k,t}$ and 
	the counterfactual conditional mean outcome is identified by 
		\begin{equation*}
	E[Y^{(0)}_{i,t}|G_i=1]=1-exp(-{R}^{(0)}_{1,t}).
\end{equation*}
\end{P6}

	\subsection{Practical Considerations for Specification Choice}
	
	With sufficiently many time periods, one can identify causal effects under restrictions of the form (\ref{SCeq1}) without any constraints on the parameters. However, in-line with standard empirical practice in diff-in-diff, researchers may wish to choose one of (\ref{eq:hlevel})  or (\ref{prophaz}). In the special case in which group-specific hazard rates are exactly equal, i.e., $c=0$ in (\ref{eq:hlevel}), then both the common dynamics and proportional hazard assumptions hold.\footnote{We thank an anonymous referee for pointing this out.} One advantage of the proportional hazard specification over common dynamics is that the latter may imply negative hazard functions, yet the definition of these objects implies they are weakly positive. This is more likely to arise if the hazard rates are close to zero and differ substantially across groups. In our empirical application we present results from both specifications and obtain very similar results.
	
	Recall that in some settings, $t$ represents the time since the beginning of a spell rather than the calendar date $t(i)$. In such settings it may be important to control for the start date of the spell. In the unemployment example, there may be seasonal differences in the job-finding rate, and thus trends in the hazard rate may differ by the date at which an individual becomes unemployed.
	
	It may also be useful to control for the start date of a spell even in some settings in which $t=t(i)$. For example, suppose we wish to assess the impact of divorce-law reform on marriage duration (e.g., \cite{Friedberg1998a}, \cite{Gruber2004}, and \cite{Wolfers2006}) by comparing rates of divorce in two states, one in which a reform is enacted, and one in which it is not. Because the reform is enacted on a set date, we let $t=t(i)$. In this case, we use data only on individuals who were married at or prior to $t(i)=1$, and thus many individuals in the data will have marriages (spells) that long precede the start of the data. Trends may differ by the time since the start of a marriage and so it may be prudent to control for the date at the start of a spell.
		
	Finally, inference in  duration settings may be complicated by random right-censoring: case in which some individuals leave a study prior to reaching the absorbing state. Formally, let $C_{i,t}$ be an indicator that individual $i$ has left the study prior to $t$. Instead of $Y_i$, we observe $\tilde{Y}_{i,t}=(1-C_{i,t})Y_{i,t}$, and an indicator $c_i$ that equals $1$ if and only if an individual is censored (i.e., $C_{i,s}=1$ and $Y_{i,s}=0$ for some $s$). Censoring complicates identification and estimation of the survival  $1-E[Y_{i,t}|G_{i}=k]$. Fortunately, numerous techniques exist to adjust for random right-censoring in estimation of the survival function. In the case in which censoring is random ($C_{i,t}$ is independent of $Y_{i,t}$  given $G_i$) the Kaplan-Meier (\cite{Kaplan1958}) estimator of the survival function is standard. If censoring is not independent, other techniques are available, see for example the review in \cite{Klein2003}. One can simply use the corresponding censoring-adjusted estimates of $1-E[Y_{i,t}|G_{i}=k]$ when applying our approach.

	\section{Estimation and Inference}
	
	The identification results in the previous section motivate plug-in estimates of counterfactual mean outcomes and treatment effects. We first define estimates $\hat{R}_{k,t}$ of the negative log survivals $R_{k,t}$ as follows:
	\begin{equation}
		\hat{R}_{k,t}:=-ln(1-\bar{Y}_{k,t}).\label{Rhatdef}
	\end{equation}
	One can informally assess the the credibility of the common dynamics (proportional hazard) assumption by visually inspecting whether there is a fixed level difference (fixed ratio) between estimated time-average hazards $\Delta_{t-1} \hat{R}_{1,t}/(t-1)$ and $\Delta_{t-1} \hat{R}_{2,t}/(t-1)$. 

	Theorem 1 motivates analogue estimates of the ATTs. 
	Under the common dynamics assumption, we estimate $c$, a post-treatment ${R}^{(0)}_{1,t}$, and the ATT, ${\tau}_{t}$ using the formulas below, where the weights $\alpha_{t}$ are positive and $\sum_{t=2}^{t^{*}}\alpha_{t}=1$.
	\begin{align}
		\hat{c}&=\sum_{t=2}^{t^{*}}\alpha_{t}(\frac{\Delta_{t-1}\hat{R}_{1,t}}{t-1}-\frac{\Delta_{t-1}\hat{R}_{2,t}}{t-1})\label{lev1}\\
		 \hat{R}^{(0)}_{1,t}&=\hat{R}_{1,1}+\Delta_{t-1} \hat{R}_{2,t}+(t-1)\hat{c}\label{eqmid} \\
		\hat{\tau}_{t}&=\bar{Y}_{1,t}-1+exp\big(-\hat{R}^{(0)}_{1,t}\big)\label{eqlev}
	\end{align}
In our empirical application, we set $\alpha=0$ for early observations and otherwise use equal weights. More generally, they may be chosen either to (a) place greater emphasis on those periods that are closer to the intervention, or (b) minimize the asymptotic variance of the estimates.

	In the proportional hazard model, the formula for $\hat{\tau}_{t}$ is as above, but we obtain estimates for $c$ and ${R}^{(0)}_{1,t}$ by
	\begin{align}
		\hat{c}&=\frac{\sum_{t=2}^{t^{*}}\alpha_{t}\Delta_{t-1}\hat{R}_{1,t}\Delta_{t-1}\hat{R}_{2,t}}{\sum_{t=2}^{t^{*}}\alpha_{t}\Delta_{t-1}\hat{R}_{1,t}^2},\,\,\,\hat{R}^{(0)}_{1,t}=R_{1,1}+\hat{c}\Delta_{t-1}\hat{R}_{2,t}\label{prop1}
	\end{align}
	
	The estimators above are both special cases of the procedure below which allows for general linear restrictions between hazard rates and is motivated by Theorem 2. Using data from the pre-treatment periods we estimate the coefficients $\{\beta_{k}\}_{k=1}^{K}$ by regressing $\Delta\hat{R}_{1,t}$ on the differenced negative log survivals of the other groups subject to the restriction that $\{\beta_{k}\}_{k=1}^{K}\in\mathcal{B}$.
	\begin{align}
		\{\hat{\beta}_{k}\}_{k=1}^{K}&=\underset{\{\beta_{k}\}_{k=1}^{K}\in\mathcal{B}}{\arg\min}\sum_{t=1}^{t^{*}}\alpha_{t}\big(\frac{\Delta_{t-1}\hat{R}_{1,t}}{t-1}-\beta_1-\sum_{k=2}^{K}\beta_{k}\frac{\Delta_{t-1}\hat{R}_{k,t}}{t-1}\big)^{2},\label{mid2}\\
		\hat{R}^{(0)}_{1,t}&=\hat{R}_{1,1}+(t-1)\hat{\beta}_1+\sum_{k=2}^{K}\hat{\beta}_{k}\Delta_{t-1}\hat{R}_{k,t}\label{estgen}
	\end{align}
	The constraint set $\mathcal{B}$ can incorporate restrictions like positivity of the weights, or that the intercept $\beta_1$ is equal to zero.

	\subsection{Estimation with Covariate Adjustment}
	
	Below we present estimators based on the covariate balancing weights in Section 2.2.1. In Appendix A.4 we describe an estimator that uses an alternative semi-parametric approach. 
	
	In order to estimate treatment effects using the weighting scheme introduced in Section 2.2.1, one must first estimate the weighting function $\omega_k$. We suggest two alternatives. The first estimator is based on (\ref{weight1}) and is appropriate for discrete covariates. In this case we simply replace the conditional probabilities in the formula (\ref{weight1}) with empirical frequencies. Let $n_k$ be the number of individuals in the sample with $G_i=k$. The estimate $\hat{\omega}_k(x)$ is defined for all $x$ for which there exists some individual $i$ in Group $k$ for whom $X_{i}=x$ and $Y_{i,1}=0$, and is given by
	\begin{equation}
		\hat{\omega}_k(x)=\frac{n_k(1-\bar{Y}_{k,1})\sum_{i=1}^{n}1\{G_{i}=1\}1\{X_{i}=x\}(1-Y_{i,1})}{n_1(1-\bar{Y}_{1,1})\sum_{i=1}^{n}1\{G_{i}=k\}1\{X_{i}=x\}(1-Y_{i,1})}\label{weight3}
	\end{equation}
	
	If some covariates are continuous or discrete with many support points, then we suggest an approach based on the propensity score  formulation (\ref{weight2}). Let $\hat{p}_k(x)$ be an estimate of $p_k(x)$. One could obtain such an estimate using say, logistic regression of $1\{G_i=k\}$ on $X_{i}$ using the sub-sample of individuals for whom $Y_{i,1}=0$. We can then estimate the weights as follows:
	\begin{equation*}
		\hat{\omega}_k(x)=\frac{n_{k}\hat{p}_1(x)(1-\bar{Y}_{k,1})}{n_{1}\hat{p}_k(x)(1-\bar{Y}_{1,1})}
	\end{equation*}
	
	Having obtained an estimate of the weight function, one can simply use the estimate below of the weighted negative log survival in place of the estimate (\ref{Rhatdef}) in the formulas (\ref{lev1}), (\ref{eqmid}), (\ref{eqlev}), (\ref{prop1}), (\ref{mid2}), or (\ref{estgen}).
	\[
	\hat{R}_{k,t}:=-ln\big(\frac{1}{n_k}\sum_{i=1}^n 1\{G_i=k\}\hat{\omega}_k(X_{i})(1-{Y}_{i,t})\big).
	\]

	\subsection{Bootstrap Inference and Specification Testing}
	
	\cite{Bertrand2004} propose the block bootstrap (\cite{Efron}) for conducting inference in vanilla diff-in-diff. The asymptotic validity of the procedure rests on an assumption that the outcome histories of different individuals are independent. However, the outcomes of a given individual may exhibit arbitrary dependence over time. We suggest the use of the block bootstrap for inference in the duration settings in this paper. Implementation follows the same steps as for standard diff-in-diff with the distinction that the bootstrap samples are used to construct our estimator of the treatment effect rather than the usual diff-in-diff estimator. Details are provided in Section B of the supplementary appendix.

	The conditions (\ref{eq:hlevel}) and (\ref{prophaz}) each have testable implications. Consider the common dynamics assumption (\ref{eq:hlevel}): if the assumption holds in all periods, then the difference in pre-treatment first-differenced negative log survivals must be constant. This motivates a test analogous to the test for parallel trends in standard difference in differences. For simplicity we focus on the common dynamics assumption, however the approach generalizes straightforwardly to the proportional hazard assumption.
	
	To motivate the test, note that under common dynamics and Assumptions 1 and 2, in all periods $t=2,...,t^*$ we have:
	\[
	\delta_t=(\frac{\Delta_{t-1}{R}_{1,t}}{t-1}-\frac{\Delta_{t-1}{R}_{2,t}}{t-1})-(\frac{\Delta_{t-1}{R}_{1,t^*}}{t-1}-\frac{\Delta_{t-1}{R}_{2,t^*}}{t-1})=0.
	\] 
	
	We use this restriction to motivate a test of the null that parallel trends holds in the pre-treatment period. In order to estimate $\delta_t$ we replace ${R}_{k,t}$ with the corresponding estimate $\hat{R}_{k,t}$. We thus obtain an estimate $\hat{\delta}_t$:
	\[
\hat{\delta}_t=(\frac{\Delta_{t-1}\hat{R}_{1,t}}{t-1}-\frac{\Delta_{t-1}\hat{R}_{2,t}}{t-1})-(\frac{\Delta_{t-1}\hat{R}_{1,t^*}}{t-1}-\frac{\Delta_{t-1}\hat{R}_{2,t^*}}{t-1}).
\] 
We then construct uniform confidence bands for $\delta_t$ using the block-bootstrap. The test rejects if and only if there is some time $2\leq t\leq t^*$ for which these confidence bands do not contain zero. Note that if covariate adjustment is used then the weight estimation or non-linear regression must also be evaluated for each bootstrap sample. The procedure is detailed in  Section B of the supplementary appendix.

Analogously, under the proportional hazards case, we have that for each $t=2,...,t^*$,	\[
\delta_t=\frac{\Delta_{t-1}{R}_{1,t}}{\Delta_{t-1}{R}_{2,t}}-\frac{\Delta_{t-1}{R}_{1,t^*}}{\Delta_{t-1}{R}_{2,t^*}}=0.
\] 
Again, one can estimate $\delta_t$ by replacing the negative log survivals with estimates and form uniform confidence bands for $\delta_t$ using the block bootstrap. A test for the null of proportional hazards then rejects if and only if there is a pre-treatment time $t$ in which the uniform confidence bands for $\delta_t$ do not contain zero.

	
	\subsection{Asymptotic Validity}
	
	All of the procedures we have described can be written as generalized method of moments estimators or sequential generalized method of moments estimators. Thus, as the sample size grows (with $K$ and $T$ fixed) standard regularity conditions ensure the consistency of our estimates and asymptotically correct coverage of the bootstrap confidence intervals. These results are well-known and so, following the example of \cite{Wooldridge2023}, we omit a formal statement here.
	
	Nonetheless, it is important to note three caveats. First, standard inferential results require identification of the nuisance parameters, which means that there must be unique coefficients $\{\beta_{k}\}_{k=1}^K$ that satisfy  (\ref{coeffid}) subject to any priori constraints that $\{\beta_{k}\}_{k=1}^K\in\mathcal{B}$. Second, if this constraint is an inequality constraint and it binds, then treatment effect estimates may not be differentiable functions of mean outcomes. In this case, asymptotic normality generally fails and the bootstrap does not have correct coverage. Such settings may call for alternative inference procedures of the form described in \cite{Fang2019}. Finally, for the estimation with covariate balancing, regular estimation generally requires that the weights $\omega_k(X_i)$ be bounded above, in which case the propensity scores must be bounded below away from zero (see \cite{Khan2010}).

	\section{Application: The Impact of Unemployment Insurance}
	
	We apply our methods to examine the impact of a policy change in Austria on the 1st of August 1989 that increased the generosity of unemployment benefits of eligible individuals. This setting was previously examined by \cite{LALIVE2006} and we use the data accompanying their paper. \cite{LALIVE2006} use the data to estimate a piece-wise constant proportional hazards model with a linear index that interacts eligibility for various aspects of the policy with an indicator that the policy change has occurred. We instead employ a cross-cohort study using our methods.  
	
	
	Following the reform, individuals aged 40-49 who had been employed for at least $312$ weeks out of the previous ten years became eligible for $39$ weeks ($273$ days) of benefits rather than the previous $30$ weeks ($210$ days). This extension applied to individuals who were already unemployed prior to the reform date.  Some individuals also qualify for a (modest) increase in the replacement rate, which is the proportion of expected earnings given to individuals receiving benefits. To simplify our analysis we consider only the increase in the potential benefit duration (PBD) rather than the change in the replacement rate. 

	We select our sample of individuals from the data as follows. First, in order to analyse only the impact of the PBD extension, we include only individuals who both (a) qualify for the PBD extension, and (b) do \textbf{not} qualify for the change in replacement rate. We further refine the sample by including those individuals who fall into one of two cohorts. To define the cohorts, let $s_i$ denote the time at which the individual became unemployed \textit{relative to the date of the benefit reform}, for example if individual $i$ became unemployed ten days after the reform then $s_i=10$. The first cohort, which is our treated group, comprises those individuals with $-210< s_i\leq 92$, i.e.,  who became unemployed fewer than $210$ days prior to the reform date, and at most $92$ days after the reform date. The second cohort forms our untreated group, and consists of those with $-575< s_i\leq -273$. This is indicated in Figure 4.1. The treated group then consists of $5,311$ individuals and the untreated group $5,390$.
	
		\begin{figure}[H]
		\caption{Cohort Definitions}
		\resizebox{350pt}{60pt}{
		\begin{tikzpicture}[x=0.018cm, y=1cm]
			
			\draw[->] (-650,0) -- (170,0) node[right]{$s_i$};
			
			\draw[thick] (0,-0.1) -- (0,0.1);
			\node[above] at (0,0.12) {Reform Date};
			
			\draw[thick, blue] (-575,-0.0) -- (-273,-0.0);
			\draw[thick, blue] (-575,0.15) -- (-575,-0.15);
			\draw[thick, blue] (-273,0.15) -- (-273,-0.15);
			\node[below, blue, font=\footnotesize] at (-575,-0.15) {$-575$};
			\node[below, blue, font=\footnotesize] at (-273,-0.15) {$-273$};
			\node[below, blue] at ({(-575-273)/2},-0.15) {Untreated Cohort};
			
			\draw[thick, red] (-210,-0.0) -- (92,-0.0);
			\draw[thick, red] (-210,0.15) -- (-210,-0.15);
			\draw[thick, red] (92,0.15) -- (92,-0.15);
			\node[below, red, font=\footnotesize] at (-210,-0.15) {$-210$};
			\node[below, red, font=\footnotesize] at (92,-0.15) {$+92$};
			\node[below, red] at ({(-210+92)/2},-0.15) {Treated Cohort};
			
			\draw[decorate, decoration={brace, amplitude=5pt}]
			(-575,0.55) -- (-210,0.55)
			node[midway, above=6pt, font=\footnotesize] {$365$ days};
			
			\draw[decorate, decoration={brace, amplitude=5pt, mirror}]
			(-273,-0.75) -- (92,-0.75)
			node[midway, below=6pt, font=\footnotesize] {$365$ days};
			
		\end{tikzpicture}
	}
		\end{figure}

	To explain this construction of the treated and untreated groups, recall that the reform extended PBD from $210$ days to $273$ and the reform applied to individuals who were already unemployed prior to the date of the reform. Thus any eligible individual who became unemployed fewer than $210$ days prior to the reform date would qualify for the PBD extension and would receive uninterrupted initial benefits for up to $273$ days into their unemployment spell. Hence we take our treated group to contain only those who became unemployed no more than $210$ days prior to the reform. Conversely, the untreated cohort consists of only individuals who became unemployed at least $273$ days prior to the reform date, and who were therefore ineligible for any benefit extension. The upper limit on $s_i$ of treated individuals ($92$ days post-reform) is then $365$ days later than the upper limit for untreated individuals, and the lower limit on $s_i$ for untreated individuals is $365$ days prior to that of the treated individuals. Thus we obtain cohorts whose unemployment start dates lie in intervals separated by one year.
	
	For each individual, the time period $t$ is taken to be the number of days since the beginning of their unemployment spell. Thus for an individual $i$, the period-$t$ outcome $Y_{i,t}$ is an indicator of whether that individual had exited unemployment at or prior to $t$ days after becoming unemployed.
	
	The extension to PBD, which is our treatment of interest, affects individuals $210$ days after becoming uemployed, suggesting a treatment date $210$ days into unemployment and thus $t^*=209$. However, individuals may anticipate the extension and search for jobs less intensively before this time than they would if their PBD were due to run out at $210$ days. In order to account for these anticipation effects, we consider treatment to occur one week earlier, so that we instead let $t^*=202$. If the benefit extension also reduces the intensity of job finding before that time, this will downwardly bias our estimated treatment effects.
	
	Given the importance of seasonal variation on job search, there may be differential trends in the rates of job-finding between individuals who became unemployed on different calendar dates in a given year. Such differential trends can be problematic if the distribution of the start dates of unemployment spells differs between the two cohorts. For this reason we apply the covariate balancing method specified in (\ref{weight3}) to re-balance the distributions of days of the year on which individuals became unemployed. In order to mitigate limited overlap, we drop from the sample individuals who became unemployed on a calendar date on which fewer than two untreated individual were made unemployed (there are $40$ such dates out of $291$). We are left with $4,511$ treated individuals and $5,364$ untreated.

	\begin{figure}[H]
		\caption{Average Outcomes and Time-Average Hazards}
		
		\subfloat[Average Outcomes]{
			\includegraphics[scale=0.23]{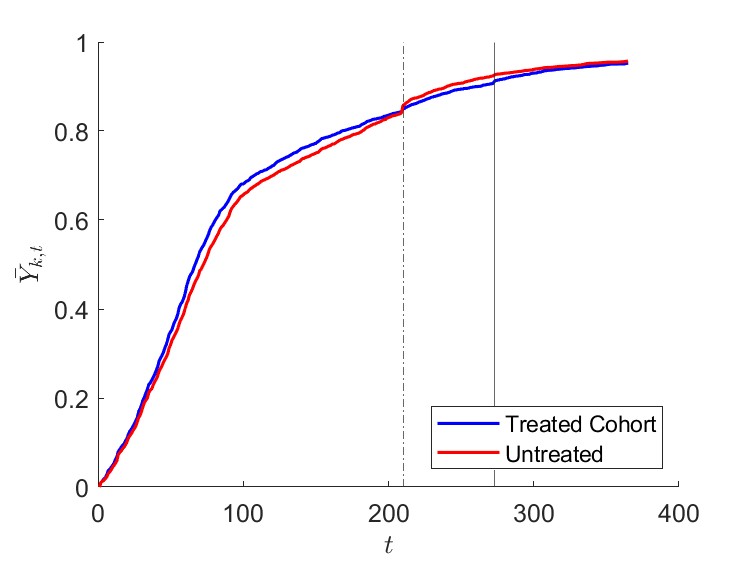}}\subfloat[Estimated Time-Average Hazards]{\includegraphics[scale=0.23]{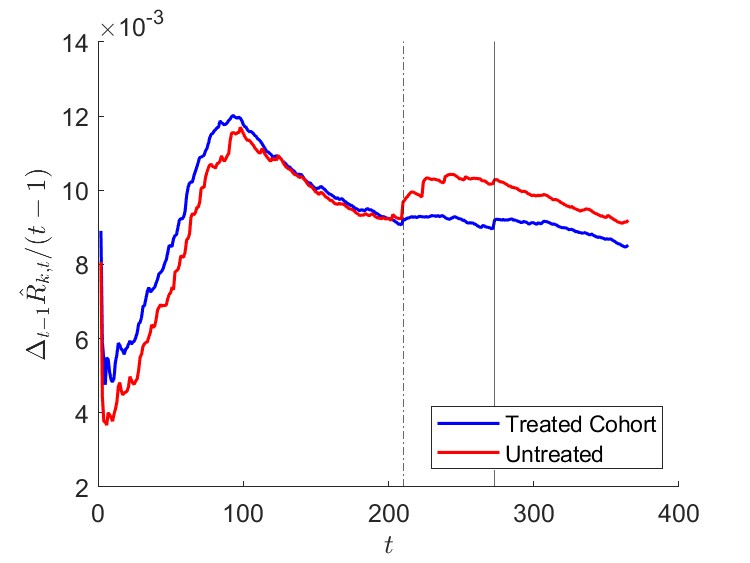}}
	\end{figure}
	Figure 4.2(a) plots average outcomes for the two groups. Recall that the mean outcome is the proportion of individuals who have left unemployment by a given day into their spell. The red curve is the mean outcome for the untreated cohort and blue for the treated cohort. The series for the two cohorts are relatively close, with a marked divergence beginning at day 210 which is marked with a vertical dashed line and is the point at which benefits expire for individuals in the untreated cohort. Over the period following the benefit extension the two curves eventually begin to converge. The rightmost vertical solid line marks day 273, at which time benefits expire for treated individuals.
	
	Figure 4.2(b) plots estimates of the time-average hazards for the two groups with covariate balancing. The curves use the same color coding as in Figure 4.2(a). There is a very marked jump in the time-average hazard rate for untreated individuals following the end of their benefit period at $210$ days into unemployment. This may suggest a sudden increase in job search intensity after unemployment benefits run out, or it may reflect a deliberate delay in the start date of new employment until exactly the date of benefit expiration.
	
	\begin{figure}[h]
		\caption{Imputations}
		
		\subfloat[Imputed Time-Average Hazards]{
			\includegraphics[scale=0.23]{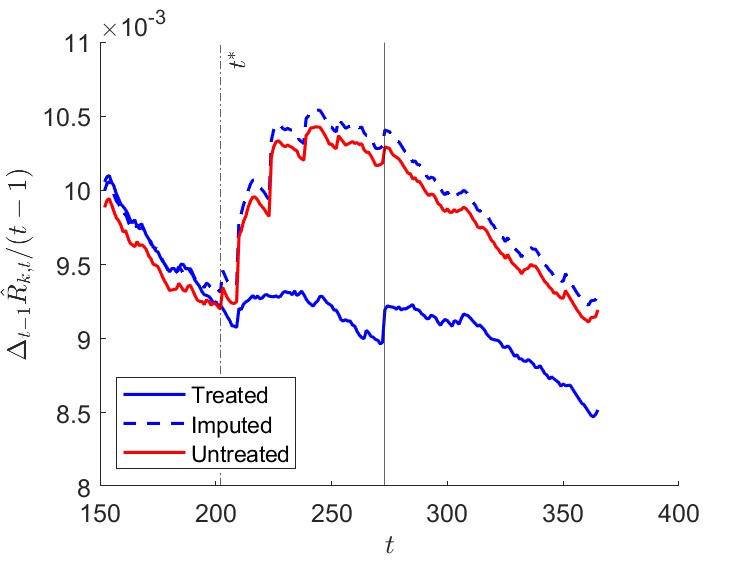}}\subfloat[Imputed Average Outcomes]{\includegraphics[scale=0.23]{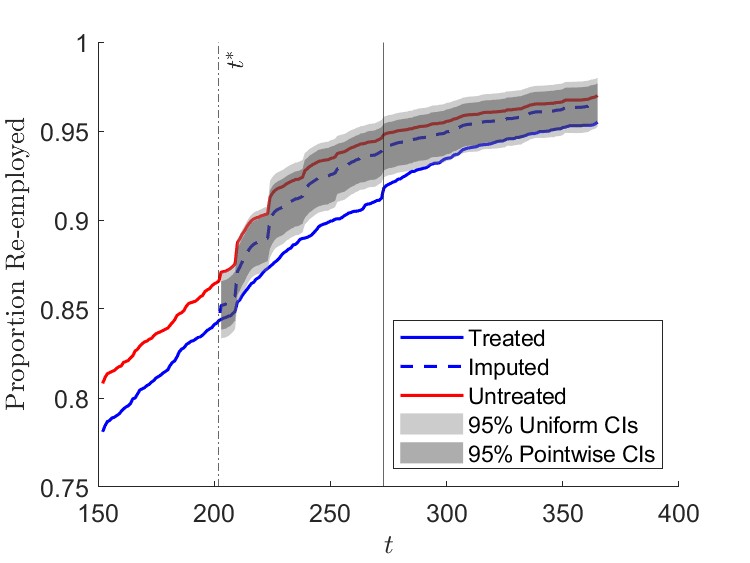}}
	\end{figure}
	
	Following standard practice we apply our diff-in-diff approach using only periods within a relatively narrow window around the treatment date. The figures in this section are for the common dynamics specification. Corresponding results from the proportional hazards specification are included in Section D of the supplementary appendix and are almost visually indistinguishable. We estimate the level difference between covariate weighted time-average hazards using only the final $50$ pre-treatment periods. This corresponds to $\alpha_t=0$ for the first $152$ periods and then $\alpha_t=\frac{1}{50}$ for the subsequent $50$ periods. This truncation is motivated by the possibility that differences in the time-average hazards sooner after unemployment may be less representative of post-treatment counterfactual differences. Expanding this window by an additional $50$ days makes little qualitative or quantitative difference.
	
	Figure 4.3(a) shows the weighted time-average hazards over this period. The dashed line indicates the imputed values for the treated cohort under the no-treatment counterfactual. Within this window, pre-treatment parallel trends appear plausible. Figure 4.3(b) plots the mean outcomes over this same window. Again, the dashed line plots the imputed mean outcomes for the treated group. The gray regions give the $95\%$ uniform and pointwise confidence bands for these imputed mean outcomes, using $10,000$ block bootstrap replications.
	
	\begin{figure}[h]
		\caption{Treatment Effects and Tests}
		
		\subfloat[Treatment Effects]{
			\includegraphics[scale=0.23]{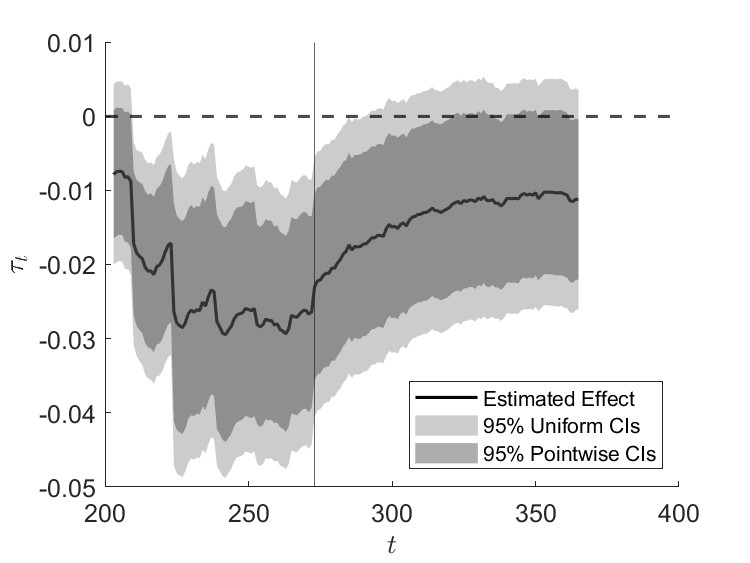}}\subfloat[Parallel Trends Test]{\includegraphics[scale=0.23]{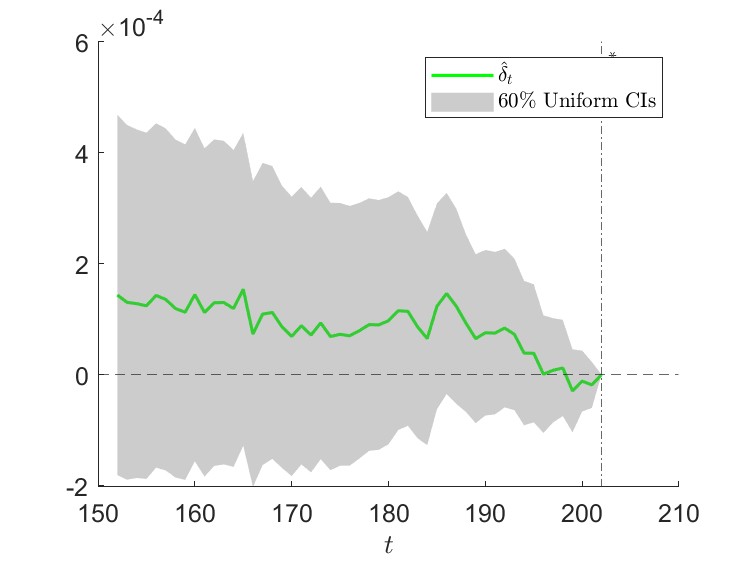}}
	\end{figure}
	
	In Figure 4.4(a) we plot the period-specific treatment effects. The treatment effect is strongly negative immediately following treatment but tends to decrease in magnitude over time. Notably, most of the steady reduction in magnitude follows the expiry of benefits for untreated individuals, which occurs at the time marked by the solid vertical line. The gray regions are $95\%$ pointwise and uniform confidence bands. The bands show a clear statistically significant negative treatment effect that builds up after the PBD extension takes effect at day $210$ before attenuating somewhat following the end of the extension at $273$ days (indicated by the solid vertical line). Note that both the pointwise and uniform confidence bands contain zero over days $203$ to $209$, and thus we do not find statistically significant evidence of anticipation of the PBD extension.
	
 Figure 4.4(b) visualizes our test for pre-treatment parallel trends. The green curve plots the difference in time-average hazard estimates relative to the value just prior to treatment. The vertical line indicates the treatment date. We see that this curve is strongly negative following treatment (in line with our negative treatment effect estimates), but is weakly positive prior to treatment. The $60\%$ uniform confidence bands contain zero over the entire pre-treatment period. This indicates a failure to reject pre-treatment parallel trends in the time-average hazards even at the highly conservative $60\%$ level (the p-value is $0.643$).

	\bibliographystyle{authordate1}
	\bibliography{hazardcites}
	
		\appendix

	\section{Additional Results}
	\subsection{Divergence of Hazard Rates Under Parallel Trends}
	Here we provide a formal argument that shows the parallel trends assumption implies divergence of counterfactual hazard rates. Suppose that $c$ is strictly positive and that for all $t$, we have  $h^{(0)}_{2}(t)>0$ and
	$E[Y _{i,t}^{(0)}|G_i=2]<1-c$. Then under parallel trends ($E[Y_{i,t}^{(0)}|G_i=1]=E[Y_{i,t}^{(0)}|G_i=2]+c$) we
	have that for any $t$,
	\[
	\frac{h^{(0)}_{1}(t)}{h^{(0)}_{2}(t)}=\frac{1-E[Y _{i,t}^{(0)}|G_i=2]}{1-E[Y _{i,t}^{(0)}|G_i=2]-c}.
	\]
	The right
	hand side grows with $t$ and it goes to infinity as the mean counterfactual outcome for Group $2$ 
	approaches the upper bound $1-c$. So as the share in Group $2$ who have reached the absorbing state approaches
	this upper bound, the ratio of the hazard rates 
	grows to infinity.
	
	A related point is made by \cite{Wu2022} who show formally that the parameter estimated by diff-in-diff must evolve over time when data are generated by a hazard model (apart from in certain degenerate cases).
	
	The analysis above is concerning because it suggests that the hazard rates exhibit aberrant behavior as 
	the proportion of individuals who have entered into an absorbing state
	grows large. The hazard rate is a primitive building block of duration
	models, and it has a clear structural interpretation. If the difference-in-differences
	assumption implies counter-intuitive and surprising behavior of the
	hazard rates, then we should reject the assumption a priori.
	
	\subsection{Identification Conditional on Covariates}
	Suppose that the a fixed linear relationship holds between counterfactual hazards within each covariate stratum:
	\begin{equation}
		h_{1}^{(0)}(t,x)=\beta_{1}(x)+\sum_{k=2}^{K}\beta_{k}(x)h_{k}^{(0)}(t,x),\label{generalCOV2}
	\end{equation}
	then the covariate-specific version of Theorem 2 below applies. This yields a conditional-on-covariates version of Theorem 1 as a    special case.
	\theoremstyle{plain} \newtheorem*{P3}{Theorem A.1} \begin{P3}
		Suppose Assumptions 1 and 2 hold. In addition suppose that (\ref{generalCOV2}) holds with $(\beta_1,\beta_2,...,\beta_K)\in\mathcal{B}$ for a known set $\mathcal{B}$.  Then $\beta_1,\beta_2,...,\beta_K$ satisfy the equations
		\begin{equation}
			\frac{\Delta_{t-1}R_{1,t}(x)}{t-1}=\beta_1+\sum_{k=2}^{K}\beta_{k}\frac{\Delta_{t-1}R_{k,t}(x)}{t-1},\,\,\forall  1<t\leq t^{*}. \label{coeffidcov}
		\end{equation}
		If the solution to the equations is unique within $\mathcal{B}$ (i.e., the equations identify $\beta_1,\beta_2,...,\beta_K$), then for $1<t\leq T$, we identify 
		$R^{(0)}_{1,t}(x)=R^{(0)}_{1,1}(x)+(t-1)\beta_1+\sum_{k=2}^{K}\beta_{k}\Delta_{t-1} R_{k,t}(x)
		$ and the counterfactual conditional mean outcome is identified by:
		\begin{equation*}
			E[Y^{(0)}_{i,t}|G_i=1,X_i=x]=1-exp\big(-R^{(0)}_{1,t}(x)\big).
		\end{equation*}
	\end{P3}
	We omit a proof as this follows immediately by an application of Theorem 2 to the subpopulation with $X_i=x$.
	
	\subsection{Staggered Adoption}
	
	We extend our result to setting with staggered adoption. In such settings,
	individuals may be treated at a range of times rather than a fixed
	time $t^{*}$. In order to accommodate such settings under our existing
	notation, we group individuals according to the time at which they
	are treated, with the group variable $G_{i}$ now referring to the
	last period before which the individual is treated. That is, if $G_{i}=k$
	then individual $i$ is treated some time after $k$ and $G_{i}=T$
	indicates that the individual is never treated within the time span
	covered by the data. In such settings, a standard approach in non-duration
	settings is to assume that the average change in the counterfactual
	outcome within Group $k$, $E[\Delta Y_{i,t}^{(0)}|G_{i}=k]$ is the
	same for all groups. Similarly, we suppose that either, for all $1\leq s\leq t\leq T$
	and $1\leq k,l\leq T$
	\begin{equation}
		h_{k}^{(0)}(t)-h_{k}^{(0)}(s)=h_{l}^{(0)}(t)-h_{l}^{(0)}(s)\label{eq:commonstagger}
	\end{equation}
	or that $h_{k}(t)$ is strictly positive for all $t$ and $k$ and
	that for all $1\leq s\leq t\leq T$ and $1\leq k,l\leq T$
	\begin{equation}
		h_{k}^{(0)}(t)/h_{k}^{(0)}(s)=h_{l}^{(0)}(t)/h_{l}^{(0)}(s).\label{eq:commonprop}
	\end{equation}
	The first condition (\ref{eq:commonstagger}) is a staggered adoption
	version of our common dynamics assumption and (\ref{eq:commonprop})
	is analogous to our proportional hazards restriction.
	
	Given the staggered adoption setting, we must update Assumption 2
	in the main text as follows.
	
	\theoremstyle{definition} \newtheorem*{A022}{Assumption $2^{*}$
		(No Anticipation/Spill-Overs: Staggered Adoption)} \begin{A022} For
		all $t\leq G_{i}$, $Y_{i,t}=Y_{i,t}^{(0)}$. \end{A022} 
	
	By similar reasoning to Section 2.1 in the main text, if Assumption
	1 in the main text holds, then under (\ref{eq:commonstagger}) for
	any $1<s\leq t\leq T$ there is a $c_{t,s}$ so that for all $k\in\{1,2,...,T\}$,
	\[
	\frac{\Delta_{t-1}R_{k,t}^{(0)}}{t-1}- \frac{\Delta_{s-1}R_{k,s}^{(0)}}{s-1}=c_{t,s}
	\]
	and under (\ref{eq:commonprop}) 
	
	\[
	\frac{\Delta_{t-1}R_{k,t}^{(0)}}{t-1}/ \frac{\Delta_{s-1}R_{k,s}^{(0)}}{s-1}=c_{t,s}.
	\]
	By Assumptions 1 and $2^{*}$, if (\ref{eq:commonstagger}) holds,
	then for any $1<s\leq t\leq T$, we have
	\begin{equation}
		\frac{\Delta_{t-1}R_{k,t}}{t-1}- \frac{\Delta_{s-1}R_{k,s}}{s-1}=c_{t,s},\,\forall k\geq t,\label{eq:esteq1}
	\end{equation}
	and if instead of (\ref{eq:commonstagger}), (\ref{eq:commonprop})
	holds, then (assuming $ \frac{\Delta_{s-1}R_{k,s}}{s-1}>0$, see discussion in Theorem 1 in
	the main text)
	\begin{equation}
		\frac{\Delta_{t-1}R_{k,t}}{t-1}/ \frac{\Delta_{s-1}R_{k,s}}{s-1}=c_{t,s},\,\forall k\geq t.\label{eq:esteq2}
	\end{equation}
	Either of these equations identifies $c_{t,s}$ for any $1<s\leq t\leq T$
	for which $P(G_{i}\geq t)>0$ . If indeed $P(G_{i}\geq t)>0$ so that
	$c_{t,s}$ is identified, then we can identify $ R_{k,t}^{(0)}$ for
	some $s\leq k<t$ by 
	\[R_{k,t}^{(0)}=R_{k,1}+(t-1)c_{t,s}+ (t-1)\frac{\Delta_{s-1}R_{k,s}}{s-1}\] in the case of
	(\ref{eq:commonstagger}) or by \[R_{k,t}^{(0)}=R_{k,t}+(t-1)c_{t,s} \frac{\Delta_{s-1}R_{k,s}}{s-1}\] in
	the case of (\ref{eq:commonprop}). Identification of group-specific
	ATTs then follows the same steps as in Theorem 1.
	
	Similar to the case in Section 2, estimation under staggered adoption
	can be carried out by the plug-in principle. Let $\mathcal{G}$ be
	the set containing all values $k\in\{1,...,T\}$ such that there is
	some individual $i$ in the sample with $G_{i}=k$. The coefficient
	$c_{t,s}$ can be estimated from the equations (\ref{eq:esteq1})
	or (\ref{eq:esteq2}) by 
	\[
	\sum_{k\in\mathcal{G}:k\geq t}\alpha_{k}( \frac{\Delta_{t-1}\hat{R}_{k,t}}{t-1}- \frac{\Delta_{s-1}\hat{R}_{k,s}}{s-1})\,\,\,\text{or}\,\,\,\sum_{k\in\mathcal{G}:k\geq t}^{T}\alpha_{k} \frac{\Delta_{t-1}\hat{R}_{k,t}}{t-1}/ \frac{\Delta_{s-1}\hat{R}_{k,s}}{s-1}
	\]
	respectively, where $\alpha_{k}$ are positive weights with $\sum_{k\in\mathcal{G}:k\geq t}\alpha_{k}=1$. Call the estimate $\hat{c}_{t,s}$.
	Then for any $k$ so that $s\leq k<t$, we can estimate $ \hat{R}_{k,t}^{(0)}=\hat{R}_{k,1}+(t-1)\hat{c}_{t,s}+ (t-1)\frac{\Delta_{s-1}\hat{R}_{k,s}}{s-1}$
	under (\ref{eq:commonstagger}) or by $ \hat{R}_{k,t}^{(0)}=\hat{R}_{k,1}+(t-1)\hat{c}_{t,s} \frac{\Delta_{s-1}\hat{R}_{k,s}}{s-1}$
	under (\ref{eq:commonprop}). Setting $s=k$ is a natural choice. ATT estimates may then be constructed
	as in Section 2.
	
	One important caveat to the above is that, if there are few individuals
	with $G_{i}=k$, then the time-average hazard estimate $ \frac{\Delta_{t-1}\hat{R}_{k,t}}{t-1}$
	is not only noisy in general but also biased, because it is a non-linear
	transformation of the sample mean. As such, estimates of $\hat{c}_{t,s}$
	are biased. This contrasts with standard staggered adoption estimates
	which are linear combinations of sample means and thus unbiased. In light
	of this, if there are few individuals with $G_{i}=k$ then one should
	set $\alpha_{k}$ small and alternative methods should be used if
	all groups are small (which may be likely to arise when the number of groups is close to the sample size). We leave the question of whether bias adjustments
	may be employed to improve performance of staggered adoption methods in duration
	settings as a topic for future work.
	
	\subsection{Semi-parametric Approach to Covariate Adjustment}
	
	We now describe an approach that restricts the manner in which covariates enter the covariate-conditional hazard rate. We suppose that a linear restriction holds between covariate-conditional hazard rates
	\begin{equation}
		h_{1}^{(0)}(t,x)=\beta_{1}(x)+\sum_{k=2}^{K}\beta_{k}(x)h_{k}^{(0)}(t,x).\label{generalCOV}
	\end{equation}
	In addition, suppose the counterfactual covariate-specific hazard within each group has the
	multiplicative form 
	\begin{equation}
		h^{(0)}_{k}(t,x)=\phi(x'\gamma_{k})\bar{h}^{(0)}_{k}(t),\label{eq:parmod}
	\end{equation}
	where $\phi(\cdot)$ is a known strictly positive link function (e.g.,
	exponential), $\gamma_{k}$ are some group-$k$-specific coefficients,
	and $\bar{h}^{(0)}_{k}(t)$ is the group-$k$-specific baseline hazard.\footnote{The baseline hazard $\bar{h}^{(0)}_{k}(t)$ differs in general from the unconditional hazard $h^{(0)}_k(t)$.} We treat the baseline hazard non-parametrically thus obtaining a semi-parametric model. The model above is a group-$k$ specific proportional hazards model (\cite{Cox1972}).
	
	In addition to (\ref{eq:parmod}), we suppose that the common dynamics, proportional hazard, or other linear restrictions holds for the \textbf{baseline
		hazards}. That is, there are coefficients
	$\beta_{1},\beta_{2},...,\beta_{K}$ so that
	\begin{equation}
		\bar{h}^{(0)}_{1}(t)=\beta_{1}+\sum_{k=2}^{K}\beta_{k}\bar{h}^{(0)}_{k}(t).\label{baselinerest}
	\end{equation}
	Given (\ref{eq:parmod}) and the restriction above, the following special case of (\ref{generalCOV}) holds
	\[
	h_{1}^{(0)}(t,x)=\phi(x'\gamma_{1})\beta_{1}+\sum_{k=2}^{K}\beta_{k}\frac{\phi(x'\gamma_{1})}{\phi(x'\gamma_{k})}h^{(0)}_{k}(t,x).
	\]
	In the $K=2$ case, imposing that $\beta_{2}=1$ and $\gamma_{1}=\gamma_{2}=\gamma$
	yields the conditional common dynamics condition $h_{1}^{(0)}(t,x)=h_{2}^{(0)}(t,x)+c(x)$ 
	with $c(x)=\phi(x'\gamma)\beta_{1}$. In the $K=2$ case with $\beta_{1}=0$
	we obtain the conditional proportional hazards restriction $h_{1}^{(0)}(t,x)=c(x)h_{2}^{(0)}(t,x)$
	with $c(x)=\beta_{2}{\phi(x'\gamma_{1})}/{\phi(x'\gamma_{2})}$.

	Under (\ref{eq:parmod}), the quantity $\Delta_{t-1}R_{k,t}^{(0)}(x)/\phi(x'\gamma_{k})$ does not depend on $x$.\footnote{In particular, under  (\ref{eq:parmod}), $\frac{\Delta_{s}R_{k,t}^{(0)}(x)}{\phi(x'\gamma_{k})}=\int_{t-s}^{t}\bar{h}^{(0)}_{k}(r)dr$.} As such, we can define
	\begin{align*}
		\Delta_{t-1}\bar{R}_{k,t}^{(0)}&:=\frac{\Delta_{t-1}R_{k,t}^{(0)}(x)}{\phi(x'\gamma_{k})}
	\end{align*} 
	Using Assumption 1 and the definition of ${R}_{k,t}^{(0)}(x)$, we obtain
	\begin{equation}
		E[Y_{i,t}^{(0)}|Y_{i,1}^{(0)}=0,G_{i}=k,X_{i}=x]=1-exp\big(-\phi(x'\gamma_{k})\Delta_{t-1}\bar{R}_{k,t}^{(0)}\big).\label{eq:identrcov}
	\end{equation}
	Under Assumptions 1 and 2, if $1\leq t\leq t^{*}$ and/or $k\neq1$ 
	then $Y_{i,t}^{(0)}=Y_{i,t}$, in which case
	\begin{equation}
		E[Y_{i,t}|Y_{i,1}=0,G_{i}=k,X_{i}=x]=1-exp\big(-\phi(x'\gamma_{k})\Delta_{t-1}\bar{R}_{k,t}^{(0)}\big).\label{eq:identrcovfact}
	\end{equation}
	For appropriate choices of $\phi$ and if $X_i$ has sufficiently rich support, the above identifies $\Delta_{t-1}\bar{R}_{k,t}^{(0)}$ for such values of $t$ and $k$.\footnote{For example, if $\phi$ is the exponential function and the support of $X_i$ contains the zero vector, then we have $E[Y_{i,t}|Y_{i,1}=0,G_{i}=k,X_{i}=0]=1-exp\big(-\Delta_{t-1}\bar{R}_{k,t}^{(0)}\big)$, which can be inverted to obtain $\Delta_{t-1}\bar{R}_{k,t}^{(0)}$. This demonstrates the importance of not including an intercept.} As we discuss in Section 3, the above is a parametric (but non-linear) regression model, and so one can use non-linear least squares to estimate the coefficients $\gamma_{k}$ and group-period fixed
	effects $\Delta_{t-1}\bar{R}_{k,t}^{(0)}$ for $t\leq t^{*}$ and/or $k\neq1$.
	
	Analogous to the case without covariates, $\Delta_{t-1}\bar{R}_{k,t}^{(0)}/(t-1)$ satisfies the same linear relationship as the baseline hazards. 
	Thus imputation proceeds similarly to Theorems 1 and 2 but with $\Delta_{t-1}\bar{R}_{k,t}^{(0)}$ replacing $ \Delta_{t-1}R_{k,t}^{(0)}$. Like in the case without covariates, one can visually assess whether $\Delta_{t-1}\bar{R}_{k,t}^{(0)}/(t-1)$ satisfies the relevant linear restriction in the pre-treatment period. Having imputed $\Delta_{t-1}\bar{R}_{1,t}^{(0)}$ in the post-treatment period, one can recover conditional mean counterfactual outcomes from  (\ref{eq:identrcov}). Compared to the case without covariates, the key distinction is that the values of $\Delta_{t-1}\bar{R}_{k,t}^{(0)}$ for $k>1$ or $t\leq t^*$ must be estimated (along with the coefficients $\gamma_k$) from the non-linear parametric regression model (\ref{eq:identrcovfact}), whereas in the case without covariates $ \Delta_{t-1}R_{k,t}$ could estimated by applying a logarithmic functions to sample average outcomes. 
	
	\theoremstyle{plain} \newtheorem*{P4}{Theorem 4} \begin{P4}
		Suppose Assumptions 1 and 2 hold. Suppose (\ref{eq:parmod}) holds for a strictly positive and strictly increasing function $\phi(\cdot)$ and $X_i$ has full support conditional on $Y_{i,t}=0$ and $G_i=k$ for all $k$. Suppose that (\ref{baselinerest}) holds with $(\beta_1,\beta_2,...,\beta_K)\in\mathcal{B}$ for a known set $\mathcal{B}$. Then ($\ref{eq:identrcovfact}$) holds and identifies $\Delta_{t-1}\bar{R}_{k,t}^{(0)}$ and $\gamma_k$ for all $k>1$ and $1<t\leq t^{*}$. Moreover, $\beta_1,\beta_2,...,\beta_K$ satisfy the equations
		\begin{equation}
			\frac{\Delta_{t-1}\bar{R}_{1,t}^{(0)}}{t-1}=\beta_1+\sum_{k=2}^{K}\beta_{k}\frac{\Delta_{t-1}\bar{R}_{k,t}^{(0)}}{t-1},\,\,\forall  1<t \leq  t^{*}. \label{coeffidcov33}
		\end{equation}
		If the solution to the equations is unique within $\mathcal{B}$, then $\Delta_{t-1}\bar{R}_{1,t}^{(0)}$ is identified (for $1<t\leq T$) by 
		$\Delta_{t-1}\bar{R}_{1,t}^{(0)}=(1-t)\beta_1+\sum_{k=2}^{K}\beta_{k}\Delta_{t-1}\bar{R}_{k,t}^{(0)}$ and 
		the counterfactual conditional mean outcome is identified by 
		\begin{align}
			&E[Y^{(0)}_{i,t}|G_i=1]\nonumber\\
			=&E[Y_{i,1}|G_i=1]+E\bigg[(1-Y_{i,1})\bigg(1-exp\big(-\phi(\gamma_{1}'X_i)\Delta_{t-1}\bar{R}_{1,t}^{(0)}\big)\bigg)\bigg].\label{formulasemi}
		\end{align}
	\end{P4}
	
	\subsubsection{Semiparametric Estimation}
	
	As an alternative to the covariate balancing approach, we present a method based on the assumptions and results in Section 2.2.2. Note that the model (\ref{eq:parmod}) is a proportional hazards model as in \cite{Cox1972}, and so the coefficients $\gamma_k$ may be estimated by Cox's partial likelihood approach using only data on untreated individuals in Group $k$ (and with $t\leq t^*$ if $k=1$). However, in order to avoid complications due to the presence of ties (that arise because $Y_{i,t}$ is only observed at a discrete set of times $t=1,...,T$) we suggest a non-linear least squares approach.
	
	To motivate the non-linear regression estimator, recall that under the assumptions of Theorem 4, if $k>1$ and/or $1<t\leq t^*$,
	\[E[Y_{i,t}|Y_{i,1}=0,G_{i}=k,X_{i}=x]=1-exp\big(-\phi(x'\gamma_{k})\Delta_{t-1}\bar{R}_{k,t}^{(0)}\big).\]
	Using the fact that conditional means minimize the mean squared error, the above implies that $\gamma_k$ and $\Delta_{t-1}\bar{R}_{k,t}^{(0)}$ jointly minimize
	\[ E\bigg[1\{G_i=k\}(1-Y_{i,1})\bigg(Y_{i,t}-1+exp\big(-\phi(\gamma_{k}'X_i)\Delta_{t-1}\bar{R}_{k,t}^{(0)}\big)\bigg)^2\bigg],\]
	Thus we simply suggest one minimizes a sample analogue of the above, summing over all $(k,t)$ such that $k>1$ and/or $1<t\leq t^*$. Denote the set of such pairs of $k$ and $t$ by $\mathcal{T}$, then we let $\{\Delta_{t-1}\hat{\bar{R}}_{k,t}\}_{(k,t)\in\mathcal{T}}$ and $\{\hat{\gamma}_k\}_{k=1}^K$ be the parameters $\{r_{k,t}\}_{(k,t)\in\mathcal{T}}$ and $\{{\gamma}_k\}_{k=1}^K$ that minimize the following non-linear least squares objective.
	\[\sum_{(k,t)\in\mathcal{T}}\sum_{i=1}^{n}1\{G_{i}=k\}(1-Y_{i1})\bigg(Y_{i,t}-1+exp\big(-\phi(x'\gamma_{k}) r_{k,t}\big)\bigg)^{2}\]
	Note that one may impose additional restrictions in the minimization above, for example that the coefficients $\gamma_k$ do not vary between groups, that is $\gamma_k=\gamma$ for all $k$. Having obtained the $\Delta_{t-1}\hat{\bar{R}}_{k,t}$, one can use this in place of $ \Delta_{t-1}\hat{R}_{k,t}$ in the formulas in Theorems 1 or 2 to obtain an estimate  $\Delta_{t-1}\hat{\bar{R}}_{1,t}^{(0)}$ of $\Delta_{t-1}\bar{R}_{1,t}^{(0)}$ for a post-treatment value of $t$. The ATT estimate $\hat{\tau}_t$ is then calculated by plugging $\Delta_{t-1} \hat{\bar{R}}_{1,t}^{(0)}$ and the coefficient estimates $\hat{\gamma}_1$ into the following empirical analogue of (\ref{formulasemi}):
	\begin{align*}
		\bar{Y}_{1,t}-\bar{Y}_{1,1}-\frac{1}{n_1}\sum_{n=1}^n (1-Y_{it})1\{G_i=1\} \bigg(1-exp\big(-\phi(\hat{\gamma}_{1}'X_i)\Delta_{t-1} \hat{\bar{R}}_{1,t}^{(0)}\big)\bigg).
	\end{align*}
	
		
	
	\section{Bootstrap Details}
	
	To carry out block bootstrap inference, one independently resamples individuals uniformly with replacement and forms a new sample using the complete series of outcomes and covariates for each individual resampled. For example, if individual $i$ is sampled in the $b$-th bootstrap iteration, then that bootstrap sample will contain an individual whose outcome history is $Y_{i,1},Y_{i,2},...,Y_{i,T}$.
	
	Having obtained block bootstrap samples, one may then evaluate bootstrap standard errors as well as pointwise and uniform confidence bands in the usual way. In particular, for each bootstrap sample $b=1,...,B$, one computes the estimate $\hat{\tau}_{t}$ using the bootstrap sample in place of the original data, and thus obtains a bootstrap estimate $\hat{\tau}^*_{b,t}$.
	
	The standard deviation $\hat{\sigma}_{t}$ of $\hat{\tau}^*_{b,t}$ over the bootstrap samples $b=1,...,B$ is taken as the standard error for  $\hat{\tau}_{t}$. To form pointwise confidence intervals for ${\tau}_{t}$, let $\hat{q}_{1-\alpha,t}$ be the $1-\alpha$-quantile of $|\hat{\tau}^*_{b,t}-\hat{\tau}_{t}|/\hat{\sigma}_{t}$. Then a $1-\alpha$-level confidence interval has the form below:
	\[
	CI_{1-\alpha,t}=[\hat{\tau}_t-\hat{q}_{1-\alpha}\hat{\sigma}_{t},\hat{\tau}_t+\hat{q}_{1-\alpha}\hat{\sigma}_{t}]
	\]
	
	The intervals described above are only designed to achieve correct pointwise coverage. Suppose we form confidence intervals for $\tau_t$ for each period from $t=t^{*}$ to $T$. Each of these intervals is specified so that it covers the corresponding period's average treatment effect with probability approximately $1-\alpha$. However, the probability that \textbf{every} one of these intervals contains its corresponding period's treatment effect may be much lower.
	
	In order to obtain a desired joint coverage probability, in place of the critical value $\hat{q}_{1-\alpha,t}$ defined above, we instead take $\hat{q}_{1-\alpha,t}$ to be the $1-\alpha$ quantile over $b=1,...,B$ of $\max_{t^* \leq s\leq T}|\hat{\tau}^*_{b,s}-\hat{\tau}_{s}|/\hat{\sigma}_{s}$ and otherwise form the confidence bands as above.
	
	Below we provide a formal description of our bootstrap inference procedures. Algorithm 1 provides instructions for constructing pointwise and uniform confidence bands for the period-specific treatment effects. Algorithm 2 details the parallel trends test.
	
	\begin{algorithm}[H]
		\caption{Block Bootstrap Inference}
		\label{alg1}
		
		\begin{algorithmic}[1]
			\STATE For each $t=t^*,,...,T$ evaluate the estimator $\hat{\tau}_{t}$ as in Section 3.1.
			
			\FOR{$b=1,2,...,B$}
			
			\STATE Independently draw a sequence of $n$  natural numbers uniformly from $\{1,2,...,n\}$. Denote the sequence by $\{j_b(1),j_{b}(2),...,j_{b}(n)\}$.
			
			\STATE For each  $t=t^*,...,T$  evaluate the estimator $\hat{\tau}_{t}$ using ${Y}_{t,j_b(i)}$ in place of ${Y}_{i,t}$, ${G}_{t,j_b(i)}$ in place of ${G}_{i,t}$, and ${X}_{j_b(i)}$ in place of ${X}_{i}$ wherever they appear in the formula. Call the resulting estimator $\hat{\tau}^*_{b,t}$.
			\ENDFOR
			
			\STATE Calculate bootstrap standard errors $\hat{\sigma}_{t}$ for $t=t^*+1,...,T$ as the standard deviation of the sample $\{\hat{\tau}^*_{b,t}\}_{b=1}^B$.
			
			\STATE For each $t=t^*+1,...,T$ let the pointwise level $1-\alpha$ critical value $\hat{q}_{1-\alpha,t}$ be the $1-\alpha$ quantile of $\{|\hat{\tau}^*_{b,t}-\hat{\tau}_{t}|/\hat{\sigma}_{t}\}_{b=1}^B$. For uniform critical values, instead use the $1-\alpha$ quantile of $\{\max_{t^*\leq s\leq T}|\hat{\tau}^*_{b,s}-\hat{\tau}_{s}|/\hat{\sigma}_{s}\}_{b=1}^B$ (note this does not depend on $t$).
			
			\STATE Form confidence bands by  $CI_{1-\alpha,t}=[\hat{\tau}_{t}-\hat{q}_{1-\alpha,t} \hat{\sigma}_{t},\hat{\tau}_{t}+\hat{q}_{1-\alpha,t} \hat{\sigma}_{t}]$

		\end{algorithmic}
	\end{algorithm}

	\begin{algorithm}[H]
		\caption{Specification testing}
		\label{alg2}
		
		\begin{algorithmic}[1]
			\STATE For each $t=2,...,t^*$ and $k=1,2$ evaluate the estimate $\Delta_{t-1} \hat{R}_{k,t}$ with formula given in Section 2. Using these estimates evaluate the difference-in-differences below to test the common dynamics case
			\[
			\hat{\delta}_{t}=(\frac{\Delta_{t-1}\hat{R}_{1,t}}{t-1}-\frac{\Delta_{t-1}\hat{R}_{2,t}}{t-1})-(\frac{\Delta_{t^*-1}\hat{R}_{1,t^*}}{t^*-1}-\frac{\Delta_{t^*-1}\hat{R}_{2,t^*}}{t^*-1})
			\]
			or for proportional hazards
			\[
			\hat{\delta}_{t}={\Delta_{t-1}\hat{R}_{1,t}}/{\Delta_{t-1}\hat{R}_{2,t}}-{\Delta_{t^*-1}\hat{R}_{1,t^*}}/{\Delta_{t^*-1}\hat{R}_{2,t^*}}.
			\]
			\FOR{$b=1,2,...,B$}

			\STATE Independently draw a sequence of $n$  natural numbers uniformly from $\{1,2,...,n\}$. Denote the sequence by $\{j_b(1),j_{b}(2),...,j_{b}(n)\}$.
			
			\STATE For $k=1,2$ and each  $t=2,...,t^*$  evaluate the estimator $ \Delta_{t-1}\hat{R}_{k,t}$ using ${Y}_{t,j_b(i)}$ in place of ${Y}_{i,t}$, ${G}_{t,j_b(i)}$ in place of ${G}_{i,t}$, and ${X}_{j_b(i)}$ in place of ${X}_{i}$ wherever they appear in the formula. Call the resulting estimator $\hat{R}^{*}_{b,k,t}$. Using these, evaluate the following quantity for common dynamics
			\[
			\hat{\delta}^{*}_{b,t}=(\frac{\Delta_{t-1}\hat{R}^{*}_{b,1,t}}{t-1}-\frac{\Delta_{t-1}\hat{R}^{*}_{b,2,t}}{t-1})-(\frac{\Delta_{t^*-1}\hat{R}^{*}_{b,1,t^*}}{t^*-1}-\frac{\Delta_{t^*-1}\hat{R}^{*}_{b,2,t^*}}{t^*-1})
			\]
			or for proportional hazards
			\[
			\hat{\delta}^{*}_{b,t}={\Delta_{t-1}\hat{R}^{*}_{b,1,t}}/{\Delta_{t-1}\hat{R}^{*}_{b,2,t}}-{\Delta_{t^*-1}\hat{R}^{*}_{b,1,t^*}}/{\Delta_{t^*-1}\hat{R}^{*}_{b,2,t^*}}.
			\]
			\ENDFOR
			
			\STATE Calculate bootstrap standard errors $\hat{\sigma}_{t}$ for $t=2,...,t^*-1$ as the standard deviation of the sample $\{\hat{\delta}^*_{b,t}\}_{b=1}^B$.
			
			\STATE For each $t=2,...,t^*-1$ let the pointwise level $1-\alpha$ critical value $\hat{q}_{1-\alpha}$ be the $1-\alpha$ quantile of $\{\max_{2\leq s\leq t^*-1}|\hat{\delta}^*_{b,s}-\hat{\delta}_{s}|/\hat{\sigma}_{s}\}_{b=1}^B$.
			
			\STATE Form confidence bands by  $CI_{1-\alpha,t}=[\hat{\tau}_{t}-\hat{q}_{1-\alpha}\hat{\sigma}_{t},\hat{\tau}_{t}+\hat{q}_{1-\alpha} \hat{\sigma}_{t}]$ 
			
			\STATE Reject pre-treatment parallel trends if for any $2\leq t\leq t^*-1$, the interval $CI_{1-\alpha,t}$ does not contain zero.
			
		\end{algorithmic}
	\end{algorithm}

	\section{Simulation Study}
	
	In order to assess the finite-sample performance of our procedure and to demonstrate the potential for large biases in standard diff-in-diff in duration settings, we implement a Monte Carlo simulation. We simulate observations from a data-generating process that obeys our identifying assumptions. The untreated counterfactual hazards for groups $1$ and $2$ are as follows:
	\begin{align*}
		h^{(0)}_1(t)&=\big(1+\sqrt{t/T}-\frac{1}{2}(t/T-1/2)^2+c\big)/(T-1)\\
		h^{(0)}_2(t)&=\big(1+\sqrt{t/T}-\frac{1}{2}(t/T-1/2)^2\big)/(T-1)
	\end{align*}
	
	The hazard rate for the observed outcomes of individuals in group $1$ (i.e., outcomes in the factual world in which an intervention occurs at time $t^*$)  is given below:
	\begin{align*}
		h_1(t)&=\big(1+\sqrt{t/T}-\frac{1}{2}(t/T-1/2)^2+c+\beta 1\{t\geq t^*\}\big)/(T-1)
	\end{align*}
	
	The probability that an individual who has yet to pass the exam passes between the discrete intervals $t-1$ and $t$ can be found by integrating and transforming the hazard rate as below:
	\[
	P(Y_{t+1,i}=1|Y_{i,t}=0,G_i=k)=1-exp\big(-\int_{t}^{t+1}h_{k}(s)ds\big)
	\]
	And similarly for counterfactual outcomes. Thus we can draw factual and counterfactual outcomes using the switching probabilities above. In practice we evaluate the integral on the right-hand side numerically.
	
	\begin{table}[H]
		\centering
		\caption{Simulation Parameters}
		
		\begin{tabular}{cccccc}
			$T$ & $t^{*}$ & $E[Y_{i,1}|G_i=1]$ & $E[Y_{i,1}|G_i=2]$ & $c$ & $\beta$\tabularnewline
			\hline 
			$20$ & $11$ & $0.4$ & $0.2$ & $0.5$ & $1$\tabularnewline
		\end{tabular}
	\end{table}
	
	The parameter values in our simulations are given in Table 1. With these parameters the population shares of individuals in groups $k=1,2$ with an outcome of $1$ under both the factual intervention and the no-intervention counterfactual are displayed in Figure A.1(a) which is identical to Figure 1.1 in the main body of the paper. The counterfactual and factual time-average hazards are illustrated in Figure A.1(b). We apply the simulations for various choices of the sample size $n$.
	
	\begin{figure}[H]
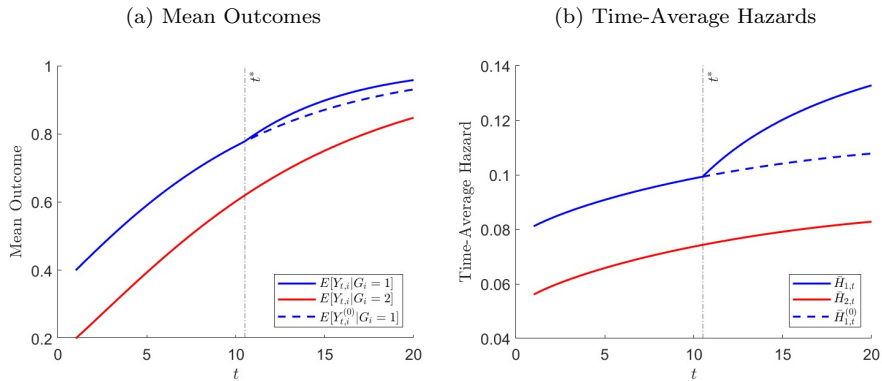

		\caption{Data Generating Process}
		
		\subfloat[Mean Outcomes]{
			
			\includegraphics[scale=0.23]{True_Mean_Outcomes.jpg}}\subfloat[Time-Average Hazards]{\includegraphics[scale=0.23]{True_Time-Average_Hazards.jpg}}
	\end{figure}

	Figure A.1(a) demonstrates that the treatment effect is positive and that the counterfactual shares for the two groups converge over time. Given our particular choice of simulation parameters, this convergence occurs primarily in the post-treatment period with trends in the pre-treatment period almost parallel. This suggests that a test for pre-treatment parallel trends in mean outcomes is unlikely to reject despite a failure of parallel trends in the post-treatment period.
	
	Figure A.1(b) shows that the counterfactual time-average hazards are parallel, which contrasts with the counterfactual mean outcomes which converge.
	
	We apply the estimation and inference procedures in Section 2. We set the weights all equal. For the block bootstrap we use $1000$ bootstrap replications.
	
	Figure A.2 contains estimates of the average treatment effects at different periods along with uniform confidence bands evaluated using the procedure in Section 2. These are from a single simulated dataset. For comparison, Figure A.3 shows standard difference-in-differences estimates and the corresponding block-bootstrap uniform confidence bands.
	
	\begin{figure}[H]
		\caption{Duration Diff-in-Diff Effect Estimates}
		
		\subfloat{\includegraphics[scale=0.155]{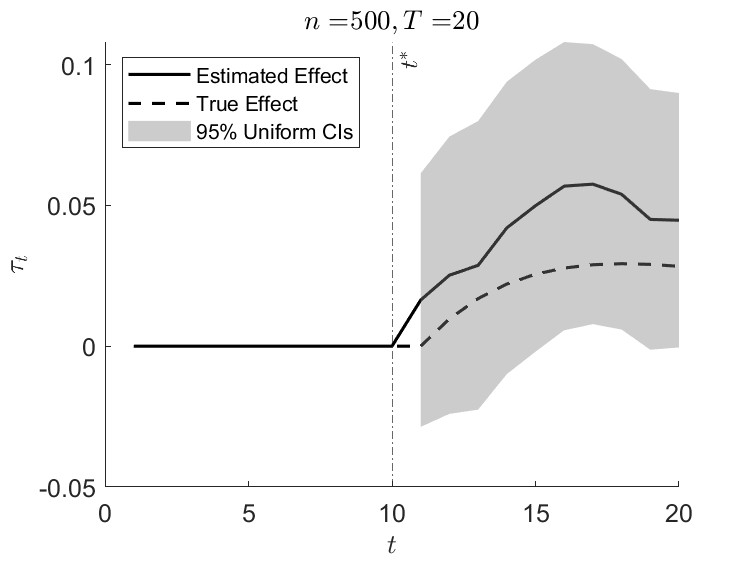}}\subfloat{\includegraphics[scale=0.155]{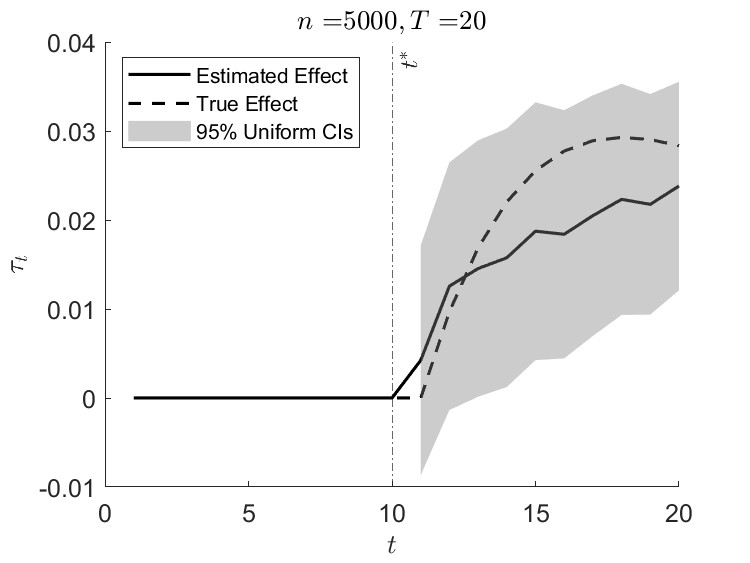}}\subfloat{\includegraphics[scale=0.155]{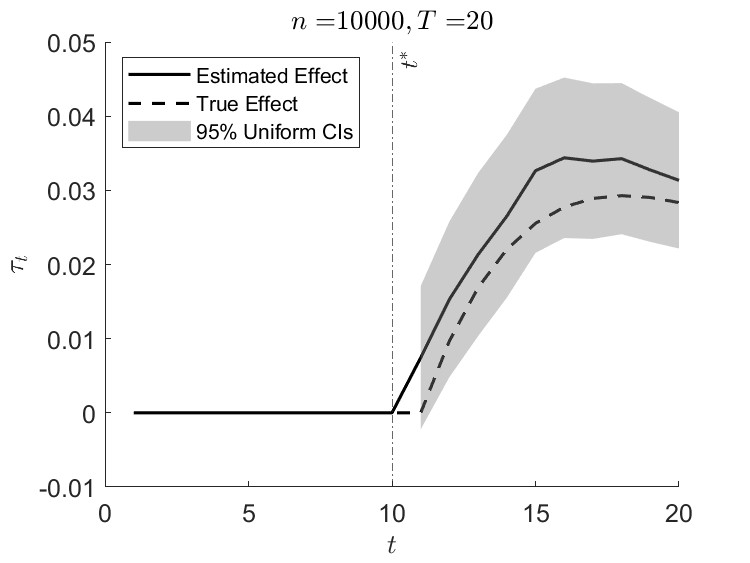}}
	\end{figure}
	
	\begin{figure}[H]
		\caption{Standard Diff-in-Diff Effect Estimates}
		
		\subfloat{\includegraphics[scale=0.155]{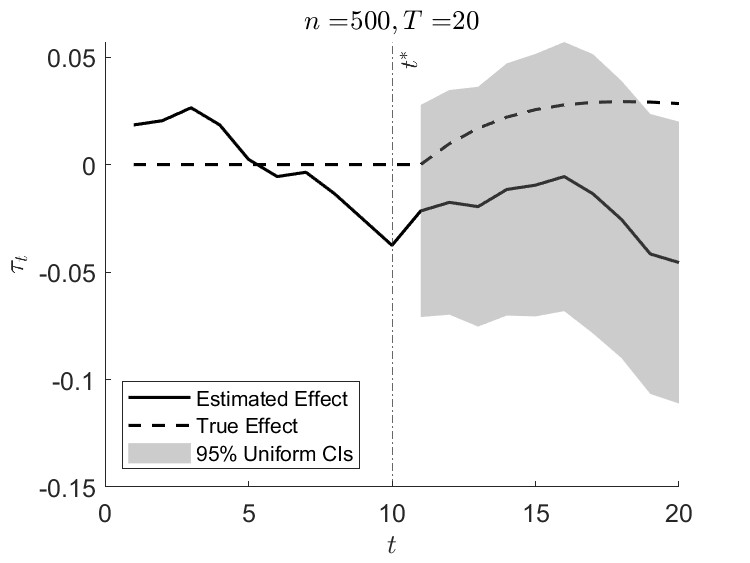}}\subfloat{\includegraphics[scale=0.155]{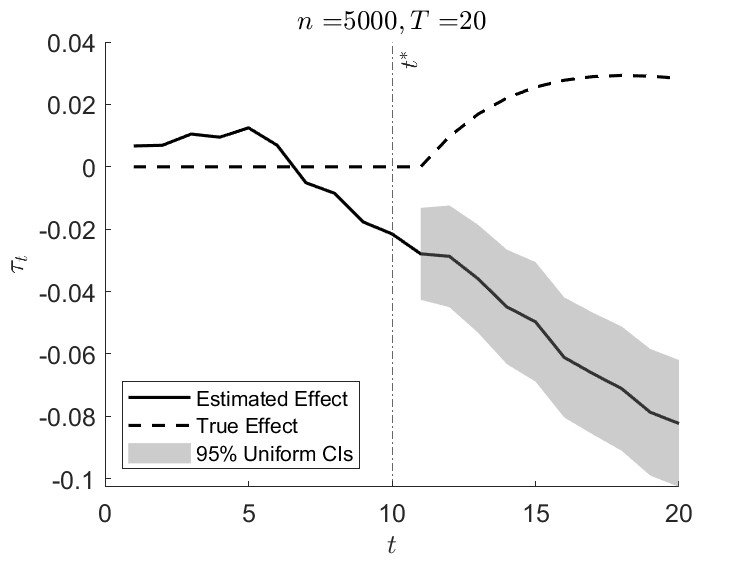}}
		\subfloat{\includegraphics[scale=0.155]{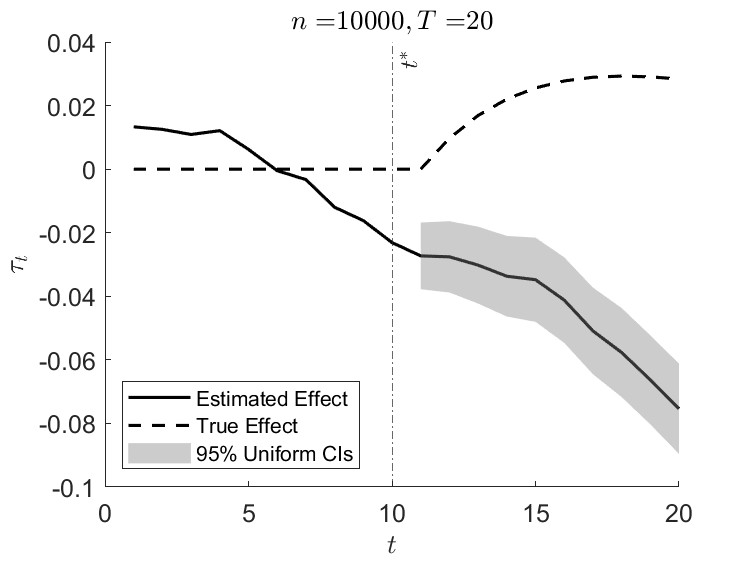}}
	\end{figure}
	
	While the estimates in Figures A.2 and A.3 are from a single simulated dataset, it is of note that even in large samples the standard diff-in-diff treatment effect estimates appear to be biased downwards. This is not surprising. In the absence of the intervention, Figure 1.a shows that the difference between group $2$ and group $1$ mean outcomes shrinks over the post-intervention period. Thus standard diff-in-diff will tend to underestimate the post-treatment difference between the mean outcomes and thus underestimate the average treatment effect. By contrast the estimates based on our method appear to be consistent.
	
	Figure A.4 contains estimates, each from a single Monte Carlo simulated dataset, of the time-average hazards for various choices of the sample size $n$. Immediately below we plot the corresponding estimates $\hat{\delta}_t$ of $\delta_t$ and associated uniform confidence bands. As we discuss in Section 2, one can test pre-treatment parallel trends in the hazard rates by observing whether or not these confidence bands contain zero in every period.
	
	\begin{figure}[H]
		\caption{Time-Average Hazard Estimates}
		
		\subfloat{\includegraphics[scale=0.155]{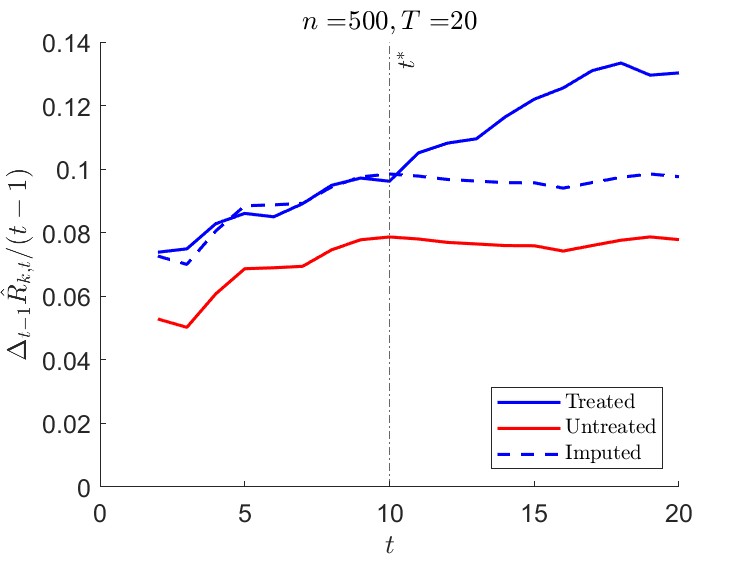}}\subfloat{\includegraphics[scale=0.155]{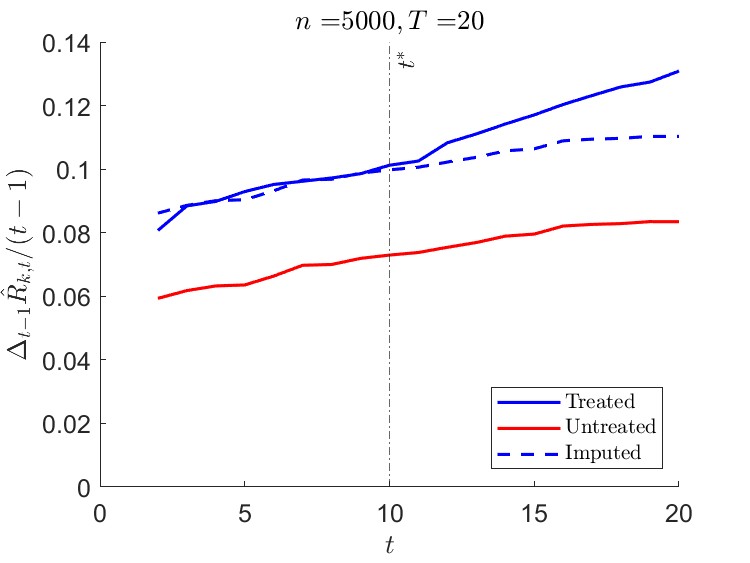}}\subfloat{\includegraphics[scale=0.155]{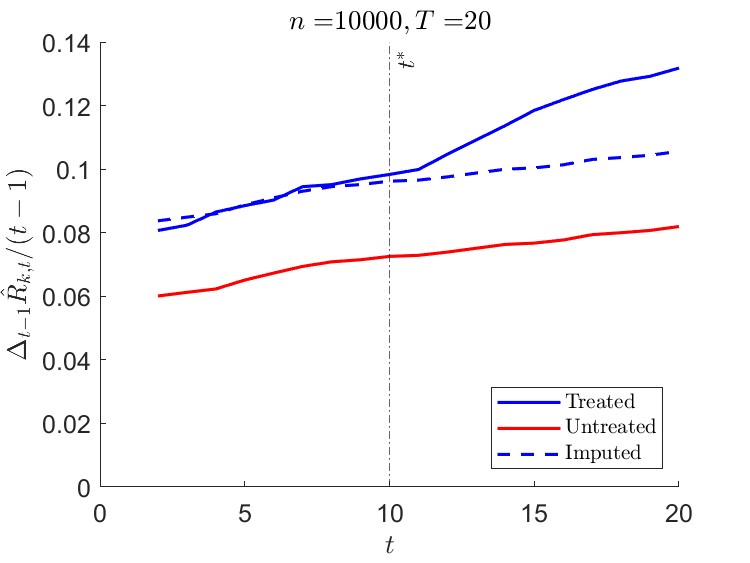}}
		
		\subfloat{\includegraphics[scale=0.155]{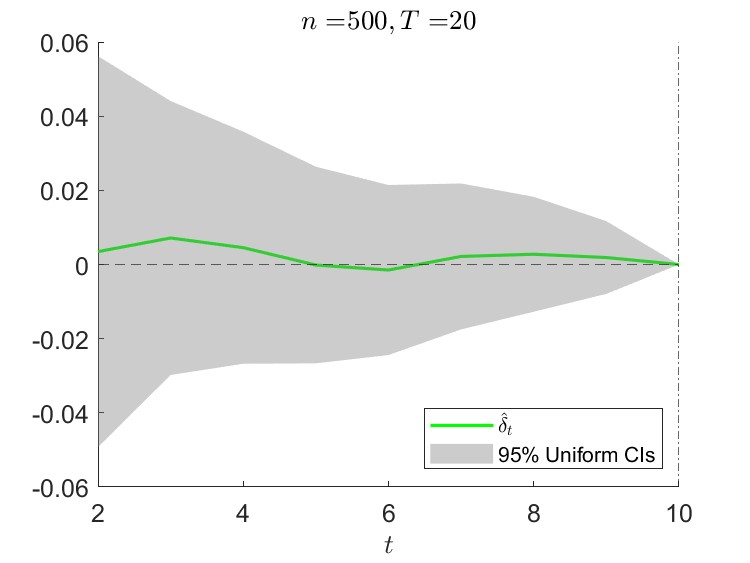}}\subfloat{\includegraphics[scale=0.155]{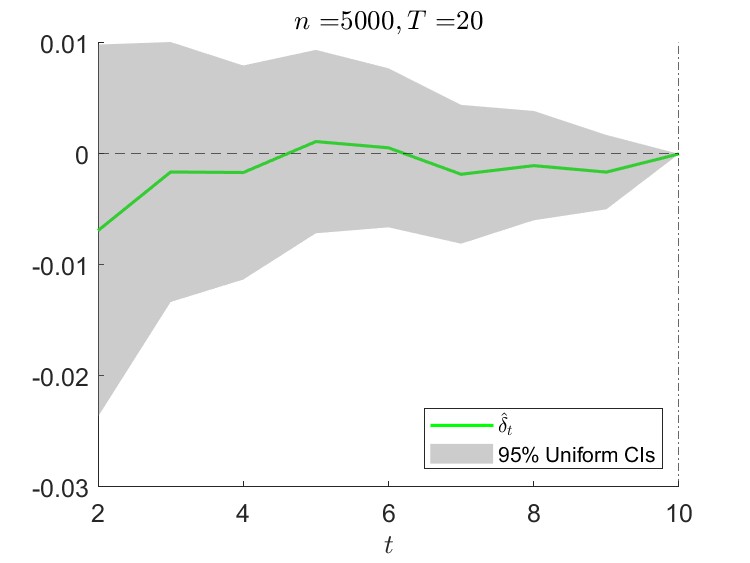}}\subfloat{\includegraphics[scale=0.155]{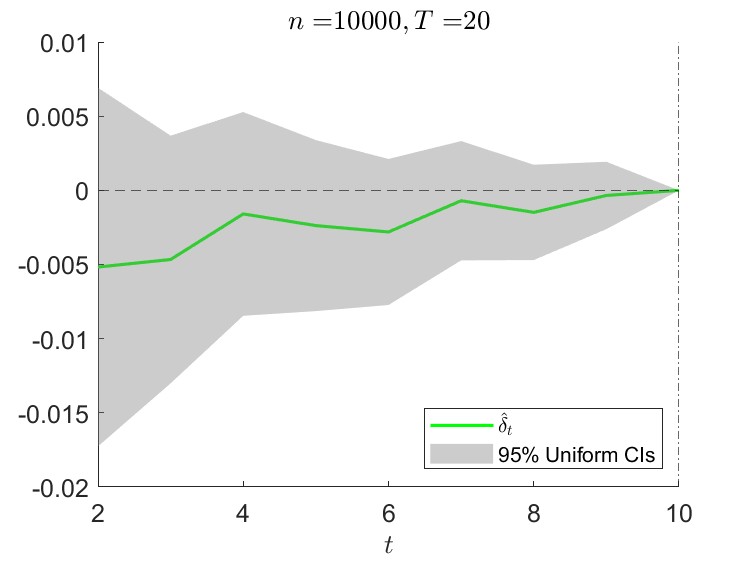}}
		
	\end{figure}

	We  simulate $1,000$ datasets. Table 2 below contains mean absolute biases, mean squared errors, and the uniform and average (over time) pointwise coverage of the uniform and pointwise confidence bands respectively. The table contains the same numbers for the standard diff-in-diff method.
	
	\begin{table}[H]
		
		\caption{Simulation Performance}
		
		\begin{tabular}{>{\centering}b{1.2cm}|>{\centering}m{1.65cm}>{\centering}m{1.65cm}>{\centering}m{1.65cm}>{\centering}m{1.75cm}>{\centering}m{1.65cm}}
			\hline 
			\multicolumn{6}{c}{Duration Difference-in-Differences}\tabularnewline
			\hline 
			\multicolumn{1}{>{\centering}b{1.2cm}}{$n$} & Absolute Bias & Mean-Squared Error & Confidence Band Coverage (Uniform) & Confidence Band Coverage (Pointwise) & Parallel Trends Test Rejects \tabularnewline
			\hline 
			$100$ & 0.00333 & 0.00176 & 0.962 & 0.957 & 0.058\tabularnewline
			$500$ & 0.00111 & 0.00034 & 0.949 & 0.951 & 0.041\tabularnewline
			$1000$ & 0.00024 & 0.00017 & 0.955 & 0.953 & 0.051\tabularnewline
			$5000$ & 0.00008 & 0.00003 & 0.945 & 0.946 & 0.058\tabularnewline
			$10000$ & 0.00010 & 0.00002 & 0.960 & 0.956 & 0.052\tabularnewline
			\hline 
			\hline 
			\multicolumn{6}{c}{Standard Difference-in-Differences}\tabularnewline
			\hline 
			\multicolumn{1}{>{\centering}b{1.2cm}}{$n$} & Absolute Bias & Mean-Squared Error & Confidence Band Coverage (Uniform) & Confidence Band Coverage (Pointwise) & Parallel Trends Test Rejects\tabularnewline
			\hline 
			$100$ & 0.070 & 0.008 & 0.680 & 0.739 & 0.038\tabularnewline
			$500$ & 0.067 & 0.006 & 0.077 & 0.266 & 0.079\tabularnewline
			$1000$ & 0.068 & 0.005 & 0.004 & 0.083 & 0.166\tabularnewline
			$5000$ & 0.068 & 0.005 & 0.000 & 0.000 & 0.576\tabularnewline
			$10000$ & 0.068 & 0.005 & 0.000 & 0.000 & 0.804\tabularnewline
			\hline 
		\end{tabular}
		
		\noindent\begin{minipage}[t]{1\columnwidth}%
			\begin{spacing}{0.5}
				{\scriptsize{}Results are from $1,000$ simulation draws. Confidence
					bands and parallel trends tests are at the $95\%$-level. Coverage of the uniform bands is uniform
					coverage, i.e., the share of simulation draws in which the bands contained
					the true treatment effects for }\textbf{\scriptsize{}all }{\scriptsize{}$t=t^{*}+1,...,T$.
					Coverage of the pointwise bands is the share of simulations in which
					the interval for period $t$ contained the true treatment effect averaged
					over}\textbf{\scriptsize{} }{\scriptsize{}$t=t^{*}+1,...,T$.
					The final column contains the share of simulated datasets in which
					the duration and standard parallel trends tests rejects, where the
					test is based on the block bootstrap and uniform confidence bands for the time-varying test statistic, as specified in Section 2.3.}
			\end{spacing}
		\end{minipage}
	\end{table}

	As one would expect, the duration diff-in-diff method, which is motivated by a correctly specified model, greatly outperforms the standard diff-in-diff procedure, which is based on a misspecified model. Encouragingly, both the uniform and pointwise confidence bands (calculated using $1000$ block bootstrap replications) appear to have approximately correct coverage. Similarly, the duration parallel trends test has approximately correct size. The standard test for parallel trends (based on the block bootstrap uniform confidence bands analogously to the duration case) does show power greater than size, that is, the test that pre-treatment trends in mean outcomes are parallel rejects with greater frequency than $5\%$ for all but the smallest sample sizes. However, the rejection frequency is fairly low apart from in large samples.
	
	\section{Empirical Application: Proportional Hazard Specification Results}
	Below we provide figures that correspond to those in Section 4 of the main paper. Here we use the proportional hazards specification rather than the common dynamics. We use the same weighted time-average hazards described in Section 4. While the figures are very close to those in Section 4, which reflects the closeness of pre-treatment the time-average hazards for Groups $1$ and $2$ so that the common dynamics and proportional hazards specifications yield similar results.

	\begin{figure}[h]
		\caption{Imputations}
		
		\subfloat[Imputed Time-Average Hazards]{
			\includegraphics[scale=0.23]{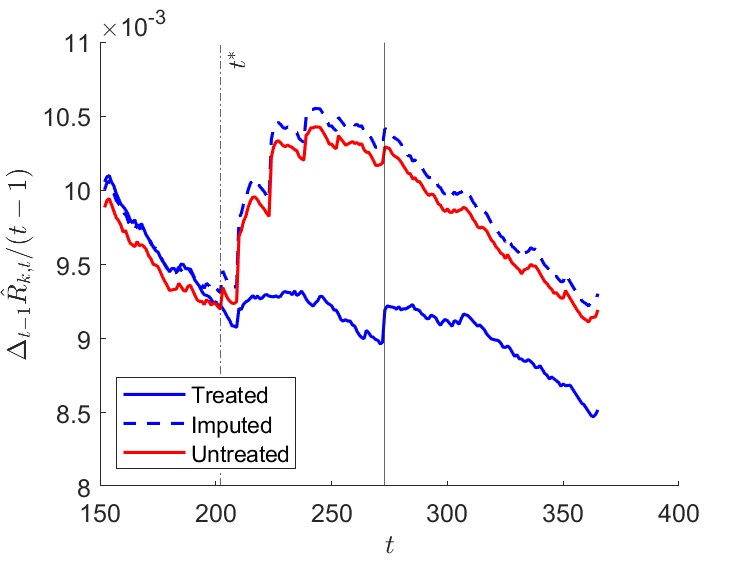}}\subfloat[Imputed Average Outcomes]{\includegraphics[scale=0.23]{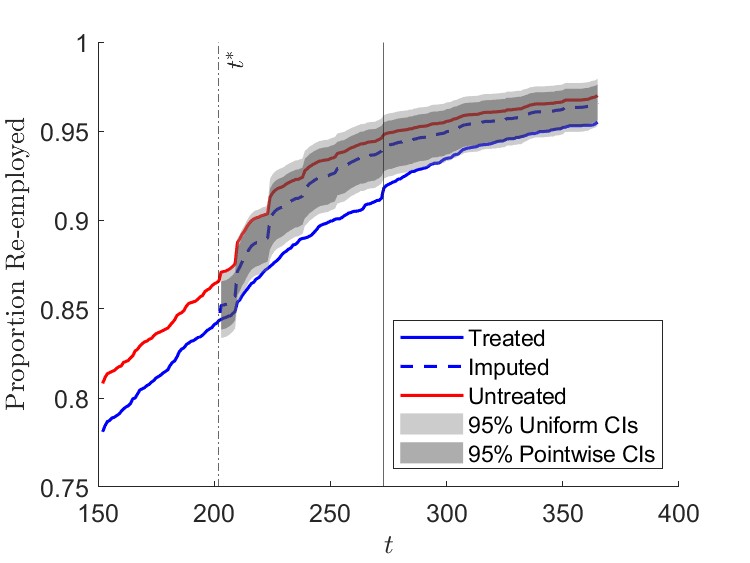}}
	\end{figure}

	\begin{figure}[h]
		\caption{Treatment Effects and Tests}
		
		\subfloat[Treatment Effects]{
			\includegraphics[scale=0.23]{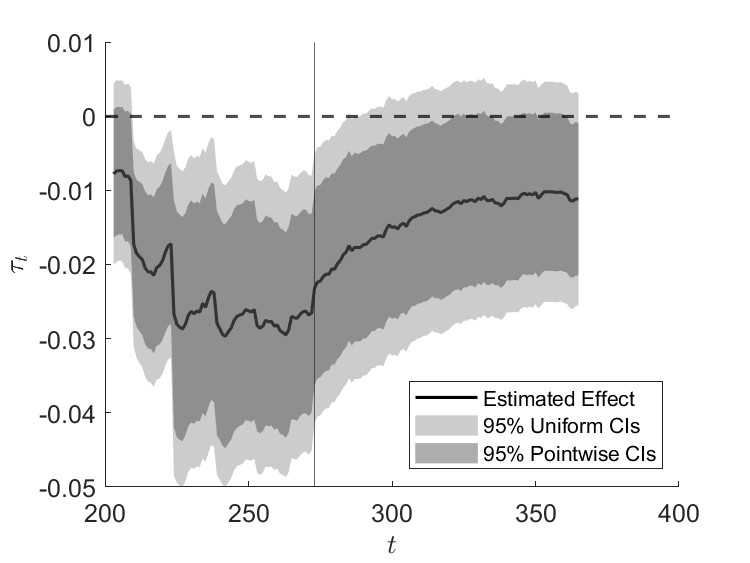}}\subfloat[Parallel Trends Test]{\includegraphics[scale=0.23]{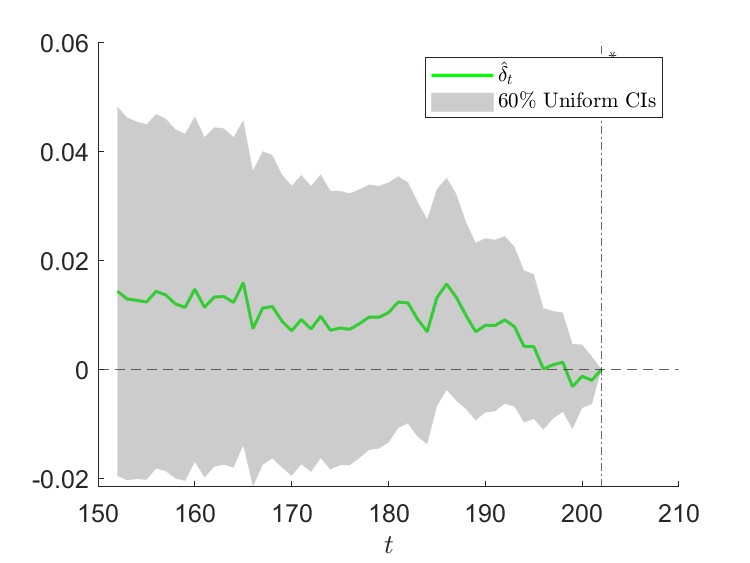}}
	\end{figure}

	\section{Proofs}
	
	\begin{proof}[Proof of Theorem 1]
		
		This result follows from Theorem 2. To see this, first note that in the case of $K=2$, the linear restriction imposed in Theorem 2 is  $h_{1}^{(0)}(t)=\beta_1+\beta_{2}h_{2}^{(0)}(t)$. Setting $\beta_1=c$ and renaming $\beta_2=1$ yields common dynamics, and setting $\beta_1=0$ and relabelling $\beta_2=c$ yields proportional hazards. We can encode the restriction $\beta_2=1$ or the restriction $\beta_1=0$ into the constraint set $\mathcal{B}$. By Theorem 2, for $1<t\leq t^*$ we obtain $\frac{\Delta_{t-1}R_{1,t}}{t-1}=\beta_1+\beta_2 \frac{\Delta_{t-1}R_{2,t}}{t-1}$ which, imposing the constraints and renaming, is equivalent to
		\[
		\frac{\Delta_{t-1}R_{1,t}}{t-1}=\frac{\Delta_{t-1}R_{2,t}}{t-1}+c\,\,\,\text{ or }\,\,\,{\Delta_{t-1}R_{1,t}}=c{\Delta_{t-1}R_{2,t}},\,\,\,\forall 1<t\leq t^{*}
		\]
		under common dynamics and proportional hazards respectively.
		
		By supposition, there are observations made for at least two time pre-treatment periods (including $t=1$). Clearly $c$ is then identified by the equations above in the common dynamics case. In the proportional hazards case, $c$ is identified if $\frac{\Delta_{t-1}R_{2,t}}{t-1}$ is non-zero for at least one value of $1<t\leq t^{*}$ (in which case $c=\frac{\Delta_{t-1}R_{1,t}}{\Delta_{t-1}R_{2,t}}$), and such a $t$ exists by supposition. Therefore $c$ is also identified in the proportional hazards case. 
		
		With $c$ identified, the remaining results follow directly from Theorem 2.
	\end{proof}
	
	\begin{proof}[Proof of Theorem 2]
		
		It is a standard result in duration analysis that the integrated hazard is equal to minus the logarithm of the survivor function. Applying this in our setting (which is valid by Assumption 1) under the no-treatment counterfactual yields
		\begin{equation*}
			R^{(0)}_{k,t}-R^{(0)}_{k,1}=\int_{1}^{t}h_{k}^{(0)}(s)ds.
		\end{equation*}
		where $R^{(0)}_{k,t}=-\ln(1-E[Y_{i,t}^{(0)}|G_{i}=k])$. Recall $ \Delta_{t-1}R_{k,t}^{(0)}:=R^{(0)}_{k,t}-R^{(0)}_{k,1}$, so for each $k$ and $t>1$
		\begin{equation}
			\frac{\Delta_{t-1}R_{k,t}^{(0)}}{t-1}=\frac{1}{t-1}\int_{1}^{t}h_{k}^{(0)}(s)ds. \label{fund}
		\end{equation}
		Taking an affine combination of both sides for different values of $k$ we obtain
		\[
		\frac{\Delta_{t-1}R^{(0)}_{1,t}}{t-1}-\sum_{k=2}^{K}\beta_{k}\frac{\Delta_{t-1}R^{(0)}_{k,t}}{t-1}= \frac{1}{t-1}\int_{1}^{t}\big(h_{1}^{(0)}(s)-\sum_{k=2}^{K}\beta_{k}h_{k}^{(0)}(s)\big)ds.
		\]
		Substituting $\beta_1=h_{1}^{(0)}(s)-\sum_{k=2}^{K}\beta_{k}h_{k}^{(0)}(s)$ which holds by supposition, and rearranging, we obtain the equality 
		\begin{equation}
			\frac{\Delta_{t-1}R^{(0)}_{1,t}}{t-1}=\beta_1+\sum_{k=2}^{K}\beta_{k}\frac{\Delta_{t-1}R^{(0)}_{k,t}}{t-1}.\label{eqH}
		\end{equation}
		Under Assumption 2.i we have that for $t\leq t^{*}$ and $k$,
		$E[Y_{i,t}^{(0)}|G_{i}=k]=E[Y_{i,t}|G_{i}=k]$, in which case  $\frac{\Delta_{t-1}R^{(0)}_{k,t}}{t-1}=\frac{\Delta_{t-1}R_{k,t}}{t-1}$ (where $\frac{\Delta_{t-1}R_{k,t}}{t-1}$ is defined identically to $\frac{\Delta_{t-1}R^{(0)}_{k,t}}{t-1}$ but with factual rather than counterfactual outcomes). Substituting gives that for all $1<t\leq t^{*}$
		\begin{equation}
			\frac{\Delta_{t-1}R_{1,t}}{t-1}=\beta_1+\sum_{k=2}^{K}\beta_{k}\frac{\Delta_{t-1}R_{k,t}}{t-1}. \label{identieqs}
		\end{equation}
		because these equations involve only functions of observables, the coefficients are identified if they are uniquely determined by the above and the a priori restriction that they are in $\mathcal{B}$. This is the first result in the theorem.
		
		By Assumption 2.ii, for all $t$ and $k\neq 1$,  $E[Y_{i,t}^{(0)}|G_{i}=k]=E[Y_{i,t}|G_{i}=k]$ and so $\frac{\Delta_{t-1}R^{(0)}_{k,t}}{t-1}=\frac{\Delta_{t-1}R_{k,t}}{t-1}$ and $R^{(0)}_{1,t}=R_{1,t}$. Substituting into  (\ref{eqH}) we see that for all $1<t\leq T$,
		\begin{equation*}
			R^{(0)}_{1,t}=R_{1,1}+(t-1)\beta_1+\sum_{k=2}^{K}\beta_{k}\Delta_{t-1}R_{k,t}.
		\end{equation*}
		which identifies $R^{(0)}_{1,t}$ if the coefficients are uniquely pinned down by (\ref{identieqs}) and the a priori restriction that they are in the set $\mathcal{B}$.
		Now, note that inverting the definition of ${R}_{1,t}^{(0)}$ we see that
		\[
		E[Y_{i,t}^{(0)}|G_i=1]=1-exp(-R_{1,t}^{(0)}).
		\]
	\end{proof}

	\begin{proof}[Proof of Proposition 1]
		Note that if $P(Y_{i,1}=0|G_{i}=k)>0$ then
		\begin{align*}
			E[\omega_{k}(X_{i})(1-Y_{i,t}^{(0)})|G_{i}=k,Y_{i,1}=0] & =\frac{E[\omega_{k}(X_{i})(1-Y_{i,t}^{(0)})(1-Y_{i,1}^{(0)})|G_{i}=k]}{P(Y_{i,1}=0|G_{i}=k)}\\
			& =\frac{E[\omega_{k}(X_{i})(1-Y_{i,t}^{(0)})|G_{i}=k]}{P(Y_{i,1}^{(0)}=0|G_{i}=k)},
		\end{align*}
		where the first equality follows by definition of conditional expectations
		and that $Y_{i,1}^{(0)}$ is binary, and the second because $(1-Y_{i,t}^{(0)})(1-Y_{i,1}^{(0)})=1-Y_{i,t}^{(0)}$
		by Assumption 1. 
		Substituting the equality above, we get
		\begin{align*}
			& E[\omega_{k}(X_{i})(1-Y_{i,t}^{(0)})|G_{i}=k]\\
			= & P(Y_{i,1}^{(0)}=0|G_{i}=k)E[\omega_{k}(X_{i})(1-Y_{i,t}^{(0)})|G_{i}=k,Y_{i,1}^{(0)}=0]\\
			= & P(Y_{i,1}^{(0)}=0|G_{i}=k)E\big[\omega_{k}(X_{i})E[1-Y_{i,t}^{(0)}|X_{i},G_{i}=k,Y_{i,1}^{(0)}]\big|G_{i}=k,Y_{i,1}^{(0)}=0\big]\\
			= & P(Y_{i,1}^{(0)}=0|G_{i}=k)E\big[E[1-Y_{i,t}^{(0)}|X_{i},G_{i}=k,Y_{i,1}^{(0)}]\big|G_{i}=1,Y_{i,1}^{(0)}=0\big],
		\end{align*}
		where the second equality follows by the law of iterated expectations,
		the third by the definition of $\omega_{k}$. Now, by the equality
		between the integrated hazard and change in log survivorship, we have
		\begin{align*}
			&ln\big(E[1-Y_{i,t}^{(0)}|X_{i},G_{i}=k,Y_{i,1}^{(0)}=0]\big) \\ =&ln\big(E[1-Y_{i,t}^{(0)}|X_{i},G_{i}=k]\big)-ln\big(E[1-Y_{i,1}^{(0)}|X_{i},G_{i}=k]\big)\\
			=&-\int_{1}^{t}h_{k}^{(0)}(s;x)ds
		\end{align*}
		and so, substituting $h_{2}^{(0)}(s;x)=h_{1}^{(0)}(s;x)-c$ and solving
		we obtain
		\begin{align*}
			E[1-Y_{i,t}^{(0)}|X_{i},G_{i}=k,Y_{i,1}^{(0)}=0] & =exp(tc)E[1-Y_{i,t}^{(0)}|X_{i},G_{i}=1,Y_{i,1}^{(0)}=0].
		\end{align*}
		Substituting this into our earlier expression for $E[\omega_{k}(X_{i})(1-Y_{i,t})|G_{i}=k]$
		(evaluated at $k=2$) and applying the law of iterated expectations
		we get
		\begin{align*}
			& E[\omega_{2}(X_{i})(1-Y_{i,t}^{(0)})|G_{i}=2]\\
			= & exp(tc)P(Y_{i,1}^{(0)}=0|G_{i}=2)E\big[E[1-Y_{i,t}^{(0)}|X_{i},G_{i}=1,Y_{i,1}^{(0)}=0]\big|G_{i}=1,Y_{i,1}^{(0)}=0\big]\\
			= & exp(tc)P(Y_{i,1}^{(0)}=0|G_{i}=2)E[1-Y_{i,t}^{(0)}|G_{i}=1,Y_{i,1}^{(0)}=0]\\
			= & exp(tc)\frac{P(Y_{i,1}^{(0)}=0|G_{i}=2)}{P(Y_{i,1}^{(0)}=0|G_{i}=1)}E[1-Y_{i,t}^{(0)}|G_{i}=1].
		\end{align*}
		
		Now note that we can rewrite the definition of $\tilde{h}_{k}^{(0)}(t)$
		as
		\begin{align*}
			\tilde{h}_{k}^{(0)}(t) & =\frac{-\frac{d}{dt}E[\omega_{k}(X_{i})(1-Y_{i,t}^{(0)})|G_{i}=k]\big)}{E[\omega_{k}(X_{i})(1-Y_{i,t}^{(0)})|G_{i}=k]\big)}\\
			& =-\frac{d}{dt}ln\big(E[\omega_{k}(X_{i})(1-Y_{i,t}^{(0)})|G_{i}=k]\big).
		\end{align*}
		
		and so
		\begin{align*}
			\tilde{h}_{2}^{(0)}(t) & =-\frac{d}{dt}ln\big(E[\omega_{2}(X_{i})(1-Y_{i,t}^{(0)})|G_{i}=2]\big)\\
			& =-c-\frac{d}{dt}ln\big(E[1-Y_{i,t}^{(0)}|G_{i}=1]\big)\\
			& =\tilde{h}_{1}^{(0)}(t)-c.
		\end{align*}
		Rearranging gives the result.
	\end{proof}
	
	\begin{proof}[Proof of Theorem 3]
		Recall from the argument in the proof of Proposition 1 that 
		\[
		\tilde{h}_{k}^{(0)}(t)=-\frac{d}{dt}ln\big(E[\omega_{k}(X_{i})(1-Y_{i,t}^{(0)})|G_{i}=k]\big).
		\]
		So integrating and using the definition $\tilde{R}_{k,t}^{(0)}:=-ln\big(E[\omega_{k}(X_{i})(1-Y_{i,t}^{(0)})|G_{i}=k]\big)$,
		we get
		\[
		\frac{\Delta_{t-1}\tilde{R}_{k,t}^{(0)}}{t-1}=\frac{1}{t-1}\int_{1}^{t}\tilde{h}_{k}^{(0)}(s)ds.
		\]
		And so, by the linear restriction on the weighted hazard functions
		we see
		\[
		\frac{\Delta_{t-1}\tilde{R}_{1,t}^{(0)}}{t-1}=\beta_{1}+\sum_{k=2}^{K}\beta_{k}\frac{\Delta_{t-1}\tilde{R}_{k,t}^{(0)}}{t-1}.
		\]
		Finally, note that trivially $\omega_{1}(X_{i})=1$ and so $\Delta_{t-1}\tilde{R}_{1,t}^{(0)}=\Delta_{t-1}R_{1,t}^{(0)}$,
		and for $k>1$, $\Delta_{t-1}\tilde{R}_{k,t}^{(0)}=\Delta_{t-1}\tilde{R}_{k,t}$ and so 
		\[
		\frac{\Delta_{t-1}R_{1,t}^{(0)}}{t-1}=\beta_{1}+\sum_{k=2}^{K}\beta_{k}\frac{\Delta_{t-1}\tilde{R}_{k,t}}{t-1}.
		\]
		The remaining steps proceed identically to Theorem 2.
	\end{proof}
	
	\begin{proof}[Proof of Theorem 4]
		By supposition $h_{k}^{(0)}(t,x)=\phi(x'\gamma_{k})\bar{h}_{k}^{(0)}(t),$
		and so using the equality between the integrated hazard and change
		in log survivorship we obtain
		\begin{align*}
			R_{k,t}^{(0)}(x)-R_{k,1}^{(0)}(x) & =\int_{1}^{t}h_{k}^{(0)}(s;x)ds\\
			& =\phi(x'\gamma_{k})\int_{1}^{t}\bar{h}_{k}^{(0)}(s)ds,
		\end{align*}
		and so 
		\begin{equation}
			\frac{\Delta_{t-1}R_{k,t}^{(0)}(x)}{\phi(x'\gamma_{k})}=\int_{1}^{t}\bar{h}_{k}^{(0)}(s)ds.\label{eq:nextbitokay}
		\end{equation}
		So the LHS does not depend on $x$. Recall the LHS is the definition of $\bar{R}_{k,t}^{(0)}$. Rearranging the definition we obtain  
		\begin{equation}
			\Delta_{t-1} R_{k,t}^{(0)}(x)=\phi(x'\gamma_{k})\Delta_{t-1}\bar{R}_{k,t}^{(0)}.\label{eq:partway2}
		\end{equation}
		Using the definition of $R_{k,t}^{(0)}(x)$, we see that if $E[1-Y_{i,1}^{(0)}|G_{i}=k,X_{i}=x]$ is non-zero, then
		\begin{align*}
			\Delta_{t-1}R_{k,t}^{(0)}(x)& =-ln\bigg(\frac{E[1-Y_{i,t}^{(0)}|G_{i}=k,X_{i}=x]}{E[1-Y_{i,1}^{(0)}|G_{i}=k,X_{i}=x]}\bigg)\\
			& =-ln(E[1-Y_{i,t}^{(0)}|G_{i}=k,X_{i}=x,Y_{i,1}^{(0)}=0])
		\end{align*}
		where the final equality follows from Assumption 1. Substituting (\ref{eq:partway2})
		into the above and solving we obtain
		\begin{equation}
			E[Y_{i,t}^{(0)}|G_{i}=k,X_{i}=x,Y_{i,1}^{(0)}=0]=1-exp\big(-\phi(x'\gamma_{k})\Delta_{t-1}\bar{R}_{k,t}^{(0)}\big).\label{eq:finstep3}
		\end{equation}
		Using Assumption 2 we obtain that for $1<t\leq t^{*}$ and/or $k\neq1$,
		\[
		E[Y_{i,t}|G_{i}=k,X_{i}=x,Y_{i,1}=0]=1-exp\big(-\phi(x'\gamma_{k})\Delta_{t-1}\bar{R}_{k,t}^{(0)}\big).
		\]
		By supposition, the support of $X_{i}$ conditional on $G_{i}$ and
		$Y_{i,1}=0$ contains $0$, and $\phi$ is strictly positive, so this
		identifies $\Delta_{t-1}\bar{R}_{k,t}^{(0)}$ for all such $t$ and $k$ so that $1<t\leq t^{*}$ and/or $k\neq1$ by
		\[
		\Delta_{t-1}\bar{R}_{k,t}^{(0)}=\frac{-ln\big(E[1-Y_{i,t}|G_{i}=k,X_{i}=0,Y_{i,1}=0]\big)}{-\phi(0)}.
		\]
		To see that $\gamma_{k}$ is also identified, note that letting $\phi^{-1}$
		denote the inverse of $\phi$ (which is strictly monotone by supposition),
		we obtain
		\[
		x'\gamma_{k}=\phi^{-1}\bigg(\frac{-ln\big(E[1-Y_{i,t}|G_{i}=k,X_{i}=x,Y_{i,1}=0]\big)}{\Delta_{t-1}\bar{R}_{k,t}^{(0)}}\bigg).
		\]
		Since $X_{i}$ has full conditional support this identifies $\gamma_{k}$.
		
		Returning to (\ref{eq:nextbitokay}) and applying the linear restriction
		on the baseline hazards, we have
		\[
		\frac{\Delta_{t-1}\bar{R}_{1,t}^{(0)}}{t-1}=\beta_{1}+\sum_{k=2}^{K}\beta_{k}\frac{\Delta_{t-1}\bar{R}_{k,t}^{(0)}}{t-1}.
		\]
		By our earlier argument, $\Delta_{t-1}\bar{R}_{k,t}^{(0)}$ is identified for
		$1<t\leq t^{*}$ and/or $k\neq1$ and so identification of $\bar{R}_{1,t}^{(0)}$
		for $t>t^{*}$ follows by the similar steps as in Theorem 2. Then
		applying (\ref{eq:finstep3}) and applying the law of iterated expectations,
		and Assumption 2.i
		\begin{align*}
			E[Y_{i,t}^{(0)}|G_{i}=1,Y_{i,1}=0] & =E\big[1-exp\big(-\phi(X_{i}'\gamma_{1})\Delta_{t-1}\bar{R}_{1,t}^{(0)}\big)\big|G_{i}=1,Y_{i,1}=0\big]\\
			& =1-\frac{E\big[(1-Y_{i,1})exp\big(-\phi(X_{i}'\gamma_{1})\Delta_{t-1}\bar{R}_{1,t}^{(0)}\big)\big|G_{i}=1\big]}{E[1-Y_{i,1}|G_{i}=1]}.
		\end{align*}
		Finally, substituting the following (which holds by Assumptions 1)
		and rearranging gives the result 
		\[
		E[Y_{i,t}^{(0)}|G_{i}=1,Y_{i,1}=0]=\frac{E[Y_{i,t}^{(0)}|G_{i}=1]-E[Y_{i,1}|G_{i}=1]}{E[1-Y_{i,1}|G_{i}=1]}.
		\]
	\end{proof}

\end{document}